\documentclass[nojss
]{jss}

\usepackage[utf8]{inputenc}

\author{
Arnald Puy\\Ecology and Evolutionary Biology,\\
Princeton University \AND Samuele Lo Piano\\School of the Built
Environment\\
University of Reading \And Andrea Saltelli\\Open Evidence Research\\
Universitat Oberta de Catalunya \And Simon A. Levin\\Ecology and
Evolutionary Biology,\\
Princeton University
}
\title{\pkg{sensobol}: an \proglang{R} package to compute variance-based
sensitivity indices}

\Plainauthor{Arnald Puy, Samuele Lo Piano, Andrea Saltelli, Simon A.
Levin}
\Plaintitle{sensobol: an R package to compute variance-based sensitivity
indices}
\Shorttitle{\pkg{sensobol}: variance-based sensitivity indices}

\Abstract{
The \proglang{R} package \pkg{sensobol} provides several functions to
conduct variance-based uncertainty and sensitivity analysis, from the
estimation of sensitivity indices to the visual representation of the
results. It implements several state-of-the-art first and total-order
estimators and allows the computation of up to third-order effects, as
well as of the approximation error, in a swift and user-friendly way.
Its flexibility makes it also appropriate for models with either a
scalar or a multivariate output. We illustrate its functionality by
conducting a variance-based sensitivity analysis of three classic
models: the \cite{Sobol1998} G function, the logistic population growth
model of \cite{Verhulst1845}, and the spruce budworm and forest model of
\cite{Ludwig1976}.
}

\Plainkeywords{R, Uncertainty, Sensitivity Analysis, Modeling}

\Address{
    Arnald Puy\\
    Ecology and Evolutionary Biology,\\
Princeton University\\
    M31 Guyot Hall\\
New Jersey 08544, USA\\
  E-mail: \email{apuy@princeton.edu}\\
  URL: \url{http://www.arnaldpuy.com}\\~\\
      Samuele Lo Piano\\
    School of the Built Environment\\
University of Reading\\
    JJ Thompson Building, Whiteknights Campus\\
Reading RG6 6AF, United Kingdom\\

      Andrea Saltelli\\
    Open Evidence Research\\
Universitat Oberta de Catalunya\\
    Carrer Almogàvers 165\\
Barcelona 08018, Spain\\

      Simon A. Levin\\
    Ecology and Evolutionary Biology,\\
Princeton University\\
    203 Eno Hall\\
New Jersey 08544, USA\\

  }

\usepackage{amsmath} \usepackage{framed} \usepackage{amsmath} \usepackage{amssymb} \usepackage{float}

\usepackage{bm} \usepackage{makecell} \usepackage{booktabs} \usepackage{tikz} \usepackage{enumitem} \usetikzlibrary{matrix,decorations.pathreplacing, calc, positioning,fit}

\begin{document}

\hypertarget{introduction}{%
\section{Introduction}\label{introduction}}

It has been argued that any form of knowledge based on mathematical
modeling is conditional on a set, perhaps a hierarchy, of either stated
or unspoken assumptions \citep{Kay2012, Saltelli2020a}. Such assumptions
range from the choice of the data and of the methods to the framing of
the problem, including normative elements that identify the nature and
the relevance of the problem itself. This conditional uncertainty is a
property of the model and not of the reality that the model has the
ambition to depict. Yet it affects the model output and hence any
model-based inference aiming at guiding policies in the ``real world''.
Identifying and understanding this conditional uncertainty is especially
paramount when the model output serves to inform a political decision,
and boils down to answering two classes of questions:

\begin{itemize} 

\item{How uncertain is the inference? Is this uncertainty compatible with the taking of a decision based on the model otcomes? Given the uncertainty, are the policy options distinguishable in their outcome?.}

\item{Which factor is dominating this uncertainty? Is this uncertainty reducible, e.g., with more data or deeper research? Are there a few dominating factors or is the uncertainty originating from several factors? Do the factors act singularly or in combination with one another?.}

\end{itemize}

The second class of questions is the realm of global sensitivity
analysis, which aims to offer a diagnosis as to the composition of the
uncertainty affecting the model output, and hence the model-based
inference \citep{Saltelli1993, Homma1996, Saltelli2008}. In helping to
appreciate the extent and the nature of the problems linked to the use
of a given model in a practical setting, global sensitivity analysis can
be considered as a tool for the hermeneutics of mathematical modeling.

Global sensitivity analysis is well represented in international
guidelines for impact assessment \citep{Azzini2020a, Gilbertson2018}, as
well as in many disciplinary journals \citep{Jakeman2006, Puy2020b}.
However, the uptake of state-of-the-art global sensitivity analysis
tools is still in its infancy. Most studies continue to prioritize local
sensitivity or one-at-a-time analyses, which explore how the model
output changes when one factor is varied and the rest is kept fixed at
their nominal values \citep{Saltelli2019a}. This approach underexplores
the input space and can not appraise interactions between factors, which
are ubiquitous in many models. Some reasons behind the scarce use of
global sensitivity analysis methods are lack of technical skills or
resources available, unawareness of global sensitivity methods or simply
reluctance due to their ``destructive honesty'': if applied properly,
the uncertainty uncovered by a global sensitivity analysis might be so
wide as to render the model largely impractical for policy-making
\citep{Leamer2010, Saltelli2019a}.

This notwithstanding, there seems to be a progressive increase in the
use of global sensitivity methods from 2005 onwards
\citep{Ferretti2016}, as well as a higher acknowledgment of them being
the ultimate acid test for the quality of any mathematical model.
Recently, global sensitivity analysis has been identified as one of the
most well-equipped scientific toolkits to tackle ``deep uncertainty''
\citep{Steinmann2020}, and a multidisciplinary team of scholars lists it
as one of the five cornerstones of responsible mathematical modeling
\citep{Saltelli2020a}.

\subsection{Sensitivity analysis packages in R and beyond}

The sparse uptake of global sensitivity methods contrasts with the many
packages available in different languages. In \proglang{Python} there is
the \pkg{SALib} package \citep{Herman2017}, which includes the Sobol',
Morris and the Fourier amplitude sensitivity test (FAST) methods. In
\proglang{MATLAB}, the \pkg{UQLab} package \citep{Marelli2014} offers
the Morris method, the \cite{Borgonovo2007} indices, Sobol' indices
(with the Sobol' and Janon estimators) and the Kucherenko indices. The
\pkg{SAFE} package \citep{Pianosi2015a}, developed originally for
\proglang{MATLAB / Octave} but with scripts available for \proglang{R}
and \proglang{Python}, includes variance-based analysis, elementary
effects and the Pianosi-Wagener method (PAWN) \citep{Pianosi2015}.

To our knowledge, there are three packages on CRAN that implement global
sensitivity analysis in \proglang{R} \citep{RCoreTeam2020}: the
\pkg{multisensi} package \citep{Bidot2018}, specifically designed for
models with a multivariate output; the \pkg{fast} package
\citep{Reusser2015}, which implements FAST; and the \pkg{sensitivity}
package \citep{Iooss2020}, the most comprehensive collection of
functions in \proglang{R} for screening, global sensitivity analysis and
robustness analysis.

\pkg{sensobol} differs from these \proglang{R} packages by the following
characteristics:

\begin{enumerate}[noitemsep]

\item \textit{It offers a state-of-the-art compilation of variance-based sensitivity estimators.} In its current version, \pkg{sensobol} comprises four first-order and eight total-order variance-based estimators, from the classic formulae of \cite{Sobol1993} or \cite{Jansen1999} to the more recent contributions by \cite{Glen2012}, \cite{Razavi2016b,Razavi2016a} (the variogram analysis of response surface total-order index, VARS-TO) or \cite{Azzini2020}. 

\item \textit{It aims at being flexible and user-friendly.} There is only one function to compute Sobol'-based sensitivity indices, \code{sobol_indices()}. Any first and total-order estimator can be simultaneously fed into the function provided that the user correctly specifies the sampling design (see Section \ref{sec:estimators}). This contrasts with the \pkg{sensitivity} package \citep{Iooss2020}, which keeps estimators compartmentalized in different functions and hence prevents the user from combining them the way it better suits their needs. Furthermore, the compatibility of \code{sobol_indices()} with the \pkg{data.table} syntax \citep{Dowle2020} makes the calculation of sensitivity indices for scalar outputs as easy as for multivariate outputs (see Section \ref{sec:budworm}).

\item \textit{It permits the computation of up to third-order effects.} Appraising high-order effects is paramount when models are non-additive (see Section \ref{sec:variance}). Although the total-order index already informs on whether a parameter is involved in interactions, sometimes a more precise account of the nature of this interaction is needed. \pkg{sensobol} opens the possibility to probe into these interactions through the computation of second and third-order effects regardless of the selected estimator.

\item \textit{It offers publication-ready figures of the model output and sensitivity-related analysis.} \pkg{sensobol} relies on \pkg{ggplot2} \citep{Wickham2016} and the grammar of graphics to yield high-quality plots which can be easily modified by the user. 

\item \textit{It is more efficient than current implementations of variance-based estimators in \proglang{R}.} Our benchmark of \pkg{sensobol} and \pkg{sensitivity} functions suggest that the former may be approximately two times faster than the latter (See Annex, Section \ref{sec:benchmark}).
\end{enumerate}

The paper is organized as follows: in Section \ref{sec:variance} we
briefly describe variance-based sensitivity analysis. In Section
\ref{sec:usage} we walk through three examples of models with different
characteristics and increasing complexity to show the main
functionalities of \pkg{sensobol}. Finally, we summarize the main
contributions of the package in Section \ref{sec:conclusions}.

\section{Variance-based sensitivity analysis}
\label{sec:variance}

Variance-based sensitivity indices use the variance to describe the
model output uncertainty. Given a model of the form \(y=f(\bm{x})\),
\(\bm{x}=(x_1,x_2,\hdots,x_i,\hdots,x_k) \in \mathbb{R}^k\), where \(y\)
is a scalar output and \(x_1,...,x_k\) are \(k\) independent uncertain
parameters described by probability distributions, the analyst might be
interested in assessing how sensitive \(y\) is to changes in \(x_i\).
One way of tackling this question is to check how much the variance in
\(y\) decreases after fixing \(x_i\) to its ``true'' value \(x_i^*\),
i.e., \(V(y|x_i=x_i^*)\). But the true value of \(x_i\) is unknown, so
instead of fixing it to an arbitrary number, we can take the mean of the
variance of \(y\) after fixing \(x_i\) to all its possible values over
its uncertainty range, while all other parameters are left to vary. This
is expressed as
\(E_{x_i} \left [V_{\bm{x}_{\sim i}}(y | x_i) \right ]\), where
\(\bm{x}_{\sim i}\) denotes all parameters-but-\(x_i\) and \(E(.)\) and
\(V(.)\) are the mean and the variance operator respectively.
\(E_{x_i} \left [V_{\bm{x}_{\sim i}}(y | x_i) \right ] \leq V(y)\), and
in fact,

\begin{equation}
V(y)=V_{x_i} \left [ E_{\bm{x}_{\sim i}}(y | x_i) \right ] + E_{x_i} \left [V_{\bm{x}_{\sim i}}(y | x_i) \right ] \,,
\end{equation}

where \(V_{x_i} \left [ E_{\bm{x}_{\sim i}}(y | x_i) \right ]\) is known
as the first-order effect of \(x_i\) and
\(E_{x_i} \left [V_{\bm{x}_{\sim i}}(y | x_i) \right ]\) is the
residual. When a parameter is important in conditioning \(V(y)\),
\(V_{x_i} \left [ E_{\bm{x}_{\sim i}}(y | x_i) \right ]\) is high.

To illustrate this property, let's imagine we run a three-dimensional
model, plot the model output \(y\) against the range of values in
\(x_i\), divide the latter in \(n\) bins and compute the mean \(y\) in
each bin. This is represented in Figure~\ref{fig:binned_mean}, with the red dots showing the
mean in each bin. The parameter whose mean \(y\) values vary the most
has the highest direct influence in the model output; in this case, it
is clearly \(x_1\). This procedure applied over very small bins is
actually \(V_{x_i} \left [ E_{\bm{x}_{\sim i}}(y | x_i) \right ]\) and
is the conditional variance of \(x_i\) on \(V(y)\), \(V_i\)
\citep{Saltelli2008}.

\begin{CodeChunk}
\begin{figure}

{\centering \includegraphics{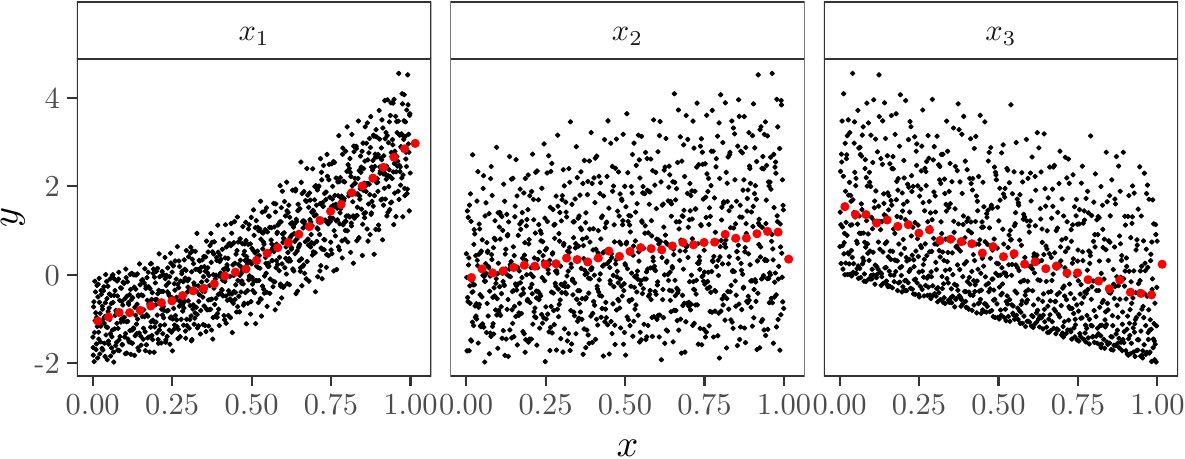} 

}

\caption{Scatterplot of $y$ against $x_i$, $i=1,2,3$. The red dots show the mean $y$ value in each bin (we have set the number of bins arbitrarely at 30), and $N=2^{10}$. The model is the polynomial function in \cite{Becker2014}, where $y=3 x_1 ^ 2 + 2 x_1 x_2 - 2 x_3$, $x_i \sim \mathcal{U}(0,1)$.}\label{fig:binned_mean}
\end{figure}
\end{CodeChunk}

When \(x_1,x_2,\hdots,x_k\) are independent parameters, \(V(y)\) can be
decomposed as the sum of all partial variances up to the \(k\)-th order,
as

\begin{equation}
V(y)=\sum_{i=1}V_i+\sum_{i}\sum_{i<j}V_{ij}+...+V_{1,2,...,k} \,,
\label{eq:decomposition}
\end{equation}

where

\begin{equation}
\begin{aligned}
V_i = V_{x_{i}}\big[E_{\bm{x}_{\sim i}}(y | x_i)\big] \hspace{4mm} 
V_{ij} &= V_{x_{i}, x_{j}}\big[E_{\bm{x}_{\sim i, j}}(y | x_i, x_j)\big] \hspace{4mm} \hdots \,. \\
& - V_{x_{i}}\big[E_{\bm{x}_{\sim i}}(y | x_i)\big] \\
& - V_{x_{j}}\big[E_{\bm{x}_{\sim j}}(y | x_j)\big]
\end{aligned}
\label{eq:Ex_i}
\end{equation}

Note that Equation~\ref{eq:decomposition} is akin to \cite{Sobol1993}'s
functional decomposition scheme:

\begin{equation}
f(\bm{x}) = f_0 + \sum_{i}f_i(x_i) + \sum_{i} \sum_{i<j}f_{ij}(x_i,x_j)+\hdots+f_{1,2,\hdots,k}(x_1,x_2,\hdots,x_k)\,,
\label{eq:functional_decomposition}
\end{equation}

where

\begin{equation}
f_0 = E(y) \hspace{8mm} 
f_i = E_{\bm{x}_{\sim _ i}} (y|x_i)-f_0 \hspace{8mm}
f_{ij} = E_{\bm{x}_{\sim _ {ij}}} (y|x_i,x_j)  - f_i - f_j - f_0 \hspace{8mm} \hdots\,,
\label{eq:Ex_i2}
\end{equation}

and therefore

\begin{equation}
V_i = V\left[f_i(x_i)\right] \hspace{8mm}
V_{ij}=V\left[f_{ij}(x_i,x_j)\right] \hspace{8mm} \hdots \,.
\label{eq:fx_ex}
\end{equation}

Function \(f(\bm{x})\) needs to be square-integrable over the dominion
of existence for the variance decomposition in Equation~\ref{eq:decomposition} to be applicable. \cite{Sobol1993} indices are
then calculated as

\begin{equation}
S_i =\frac{V_i}{V(y)} \hspace{8mm} S_{ij}=\frac{V_{ij}}{V(y)} \hspace{8mm} \hdots \,,
\label{eq:terms}
\end{equation}

where \(S_i\) is the first-order effect of \(x_i\), \(S_{ij}\) is the
second-order effect of \((x_i,x_j)\) (formed by the first order effect
of \(x_i\), \(x_j\) and their interaction), etc. \(S_i\) (\(S_{ij}\))
can thus be expressed as the fractional reduction in the variance of
\(y\) which will be obtained if \(x_i\) (\(x_i,x_j\)) could be fixed. In
variance-based sensitivity analysis, \(S_i\) is used to rank parameters
given their contribution to the model output uncertainty, a setting
known as ``factor prioritization'' \citep{Saltelli2008}.

If we divide all terms in Equation~\ref{eq:decomposition} by \(V(y)\),
we get

\begin{equation}
\sum_{i=1}^{k}S_i+\sum_{i}\sum_{i<j}S_{ij}+...+S_{1,2,...,k} = 1.
\label{eq:decomposition_summand}
\end{equation}

When \(\sum_{i=1}^kS_i=1\), the model is additive, i.e., the variance of
\(y\) can be fully decomposed as the sum of first-order effects, meaning
that there are no interaction between parameters. However, this is
rarely the case in real-life models, and first-order indices are usually
not enough to account for all the model output variance.

This is demonstrated with the example in Figure~\ref{fig:ishi_plot}: \(x_2\) and \(x_3\)
do not have a first-order effect on \(y\) as
\(V_{x_i} \left [ E_{\bm{x}_{\sim i}}(y | x_i) \right ] \approx 0\).
However, and unlike \(x_2\), \(x_3\) does influence \(y\) given the
shape of the scatterplot, so it can not be an inconsequential parameter.
Indeed, \(x_3\) influences \(y\) through high-order effects, i.e., by
interacting with some other parameter(s). In this specific case, it is
clear that \(x_3\) must interact with \(x_1\) given that \(x_2\) is
non-influential. Such appraisal of interactions can rarely be made
through the visual inspection of scatterplots alone, and often requires
computing higher-order terms in Equation~\ref{eq:decomposition_summand}.

Since there are \(2^k-1\) terms in Equation~\ref{eq:decomposition_summand}, a model with 10 parameters will have
1023 terms, making a full variance decomposition very arduous: just the
computation of second-order terms for this model would require
estimating 45 indices.

To circumvent this issue, \cite{Homma1996} proposed to compute the
total-order index \(T_i\), which measures the first-order effect of
\(x_i\) jointly with its interactions with all the other parameters. In
other words, \(T_i\) includes all terms in Equation~\ref{eq:decomposition} with the index \(i\), and is computed as follows:

\begin{equation}
T_i=1 - \frac{V_{\bm{x}_{\sim i}}\big[E_{x_i}(y | \bm{x}_{\sim i})\big]}{V(y)} = \frac{E_{\bm{x}_{\sim i}}\big[V_{x_{i}}(y | \bm{x}_{\sim i})\big]}{V(y)} \,, 
\label{eq:Ti}
\end{equation}

For a three-dimensional model, the total-order index of \(x_1\) will
thus be computed as \(T_1=S_1 + S_{1,2} + S_{1,3} + S_{1,2,3}\). Since
\(T_i=0\) indicates that \(x_i\) does not convey any uncertainty to the
model output, the total-order index has been used to screen influential
from non-influential parameters, a setting known as ``factor fixing''
\citep{Saltelli2008}.

\begin{CodeChunk}
\begin{figure}

{\centering \includegraphics{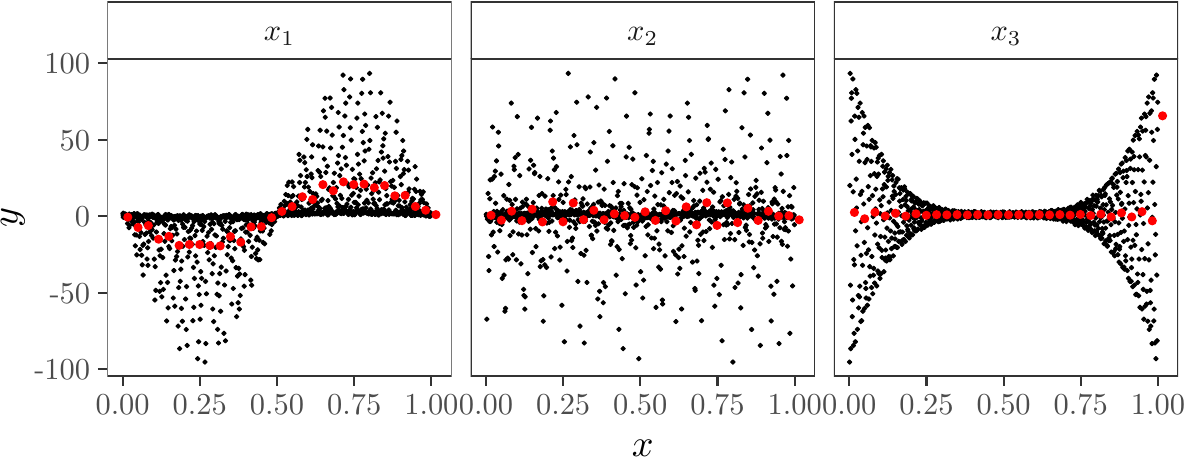} 

}

\caption{Scatterplot of $y$ against $x_i$, $i=1,2,3$. The red dots show the mean $y$ value in each bin (we have set the number of bins arbitrarely at 30), and $N=2^{10}$. The model is the \cite{Ishigami1990} function.}\label{fig:ishi_plot}
\end{figure}
\end{CodeChunk}

The popularity of variance-based methods derives from their capacity to
provide sensitivity measures that are model-independent and easily
understandable. They also capture the influence of the full range of
variation in each parameter, including its interactions with the rest
\citep{Saltelli2008}. Some known limitations are their high
computational demands and that they may not be the most appropriate
proxy of uncertainty when the output distribution is highly skewed or
multi-modal \citep{Pianosi2015}.

\subsection{Sampling design and sensitivity estimators}
\label{sec:estimators}

The computation of variance-based sensitivity indices requires two
elements: 1) a sampling design, i.e., a strategy to arrange the sample
points into the multidimensional space of the input factors, and 2) an
estimator, i.e., a formula to compute the sensitivity measures
\citep{LoPiano2021}. Both elements are intertwined: the reliance on a
given sampling design determines which estimators can be used and the
other way around.

\pkg{sensobol} v.1.0.3 currently offers support for four first-order and
eight total-order sensitivity estimators, which rely on specific
combinations of \(\bm{A}\), \(\bm{B}\), \(\bm{A}_B^{(i)}\) or
\(\bm{B}_A^{(i)}\) matrices (Tables~\ref{tab:si_estimators}--\ref{tab:ti_estimators}). Estimator 9 in Table
\ref{tab:ti_estimators} is known as VARS-TO and requires a different
sampling design based on star-centers and cross-sections
\citep{Razavi2016b, Razavi2016a}. We provide further information about
VARS-TO in the Annex, section \ref{sec:VARS}. All these estimators are
sample-based and hence \pkg{sensobol} does not include emulators or
surrogate models.

\begingroup
\renewcommand{\arraystretch}{1.7}
\begin{table}[ht]
\centering
\begin{tabular}{lllp{3.2cm}}
\toprule
Nº & Estimator & \code{first} & Author \\
\midrule
1 & $\frac{\frac{1}{N} \sum_{v=1}^{N} f(\bm{A})_v f(\bm{B}^{(i)}_A)_v - f^2_0}{V(y)}$ & \code{"sobol"} &\cite{Sobol1993} \\
2 & $\frac{\frac{1}{N}\sum_{v=1}^{N}f(\bm{B})_v \left [ f(\bm{A}^{(i)}_B)_v-f(\bm{A})_v \right ]}{V(y)}$ & \code{"saltelli"} & \cite{Saltelli2010a} \\
3 & $\frac{V(y) - \frac{1}{2N} \sum_{v=1}^{N} \left [ f(\bm{B})_v - f(\bm{A}^{(i)}_B)_v \right ] ^ 2}{V(y)}$ & \code{"jansen"} & \cite{Jansen1999} \\
4 & $\frac{2 \sum_{v=1}^{N} ( f(\bm{B}_A^{(i)})_v - f(\bm{B})_v  )   ( f(\bm{A})_v - f(\bm{A}_B^{(i)})_v  )}{\sum_{v=1}^{N} \left [ ( f(\bm{A})_v - f(\bm{B})_v  )^2 + ( f(\bm{B}_A^{(i)})_v - f(\bm{A}_B^{(i)})_v  )^2 \right ]}$ & \code{"azzini"} & \cite{Azzini2020} \\
\bottomrule
\end{tabular}
\caption{\label{tab:si_estimators}First-order estimators included in \pkg{sensobol} (v1.0.3). $f_0 = \frac{1}{2N} \sum_{v=1}^{N} \left [ f(\bm{A})_v + f(\bm{B})_v \right ]$ and $V(y) = \frac{1}{2N - 1} \sum_{v = 1}^{N} \left [ (f(\bm{A})_v - f_0)^2 + (f(\bm{B})_v - f_0)^2 \right ]$.}
\end{table}
\endgroup

\begingroup
\renewcommand{\arraystretch}{1.7}
\begin{table}[ht]
\centering
\begin{tabular}{lllp{3.2cm}}
\toprule
Nº & Estimator & \code{total} & Author \\
\midrule
1 & $\frac{\frac{1}{2N} \sum_{v=1}^{N } \left [ f(\bm{A})_v - f(\bm{A}_{B} ^{(i)})_v \right ] ^ 2}{V(y)}$ & \code{"jansen"} & \cite{Jansen1999} \\
2 & $\frac{\frac{1}{N} \sum_{v=1} ^{N} f(\bm{A})_v \left [ f(\bm{A})_v - f(\bm{A}_{B} ^{(i)})_v\right ]}{V(y)}$ & \code{"sobol"} & \cite{Sobol2001} \\
3 &  $\frac{V(y) - \frac{1}{N} \sum_{v = 1}^{N} f(\bm{A}_v)  f(\bm{A}_{B} ^{(i)})_v + f_0^2}{V(y)}$ & \code{"homma"} &\cite{Homma1996} \\
4 & $1 - \frac{\frac{1}{N} \sum_{v=1}^{N} f(\bm{B})_v f(\bm{B}^{(i)}_A)_v - f_0^2}{\frac{1}{N} \sum_{v=1}^{N} f(\bm{A})_v^2 - f_0^2}$ & \code{saltelli} & \cite{Saltelli2008} \\
5 & $1 - \frac{\frac{1}{N} \sum_{v=1}^{N}f(\bm{A})_v f(\bm{A}_{B} ^{(i)})_v - f_0^2}{\frac{1}{N} \sum_{v=1}^{N} \frac{f(\bm{A}_v)^2 + f(\bm{A}_{B} ^{(i)})_v^2}{2} - f_0^2}$ & \code{"janon"} & \cite{Janon2014} \newline \cite{Monod2006a}  \\
6 &  $ 1 - \left [\frac{1}{N-1}\sum_{v=1}^{N} \frac{\left [ f(\bm{A})_v - \left \langle f(\bm{A})_v\right \rangle \right ] \left[ f(\bm{A}_{B} ^{(i)})_v - \left \langle f(\bm{A}_{B} ^{(i)})_v\right \rangle \right ]}{\sqrt{V\left [f(\bm{A})_v\right ] V\left [f(\bm{A}_{B} ^{(i)})_v \right]}} \right ]$ & \code{"glen"} &  \cite{Glen2012} \\
7 & $\frac{\sum_{v = 1}^{N} [ f(\bm{B})_v - f(\bm{B}^{(i)}_A)_v ] ^ 2 +   [ f(\bm{A})_v - f(\bm{A}^{(i)}_B)_v  ] ^ 2}{\sum_{v = 1} ^ {N} [ f(\bm{A})_v - f(\bm{B}) _v  ] ^ 2 +  [ f(\bm{B}^{(i)}_A)_v - f(\bm{A}^{(i)}_B)_v  ] ^ 2}$ & \code{"azzini"} & \cite{Azzini2020} \\
8 & $\frac{E_{x^*_{\sim_{i}}}\left [ \gamma_{x^*{_\sim i}}(h_i)\right] + E_{x^*{_\sim i}} \left [ C_{x^*{_\sim i}}(h_i) \right ] }{V(y)}$ & \makecell{See Annex, \\ section \ref{sec:VARS}} & \cite{Razavi2016a, Razavi2016b}. \\
\bottomrule
\end{tabular}
\caption{\label{tab:ti_estimators} Total-order estimators included in \pkg{sensobol} (v1.0.3). $f_0$ and $V(y)$ are estimated according to the original papers. See Table 1 in \cite{Puyj} for a description of their calculation.}
\end{table}
\endgroup

How are these matrices formed, and why are they required? Let \(\bm{Q}\)
be a \((N, 2k)\) matrix constructed using either random or quasi-random
number generators, such as the \cite{Sobol1967, Sobol1976} sequence or a
Latin hypercube sampling design \citep{McKay1979}. The \(\bm{A}\) and
the \(\bm{B}\) matrices include respectively the leftmost and rightmost
\(k\) columns of the \(\bm{Q}\) matrix. The \(\bm{A}_B^{(i)}\)
\((\bm{B}_A^{(i)}\)) matrices are formed by all columns from the
\(\bm{A}\) \((\bm{B}\)) matrix except the \(i\)-th, which comes from
\(\bm{B}\) \((\bm{A})\) (Equation~\ref{eq:ab_matrices}, Figure~\ref{fig:ab_matrices}).

\begin{figure}[!ht]
\centering
\begin{equation}
\begin{aligned}
  \bm{Q} & = 
\begin{tikzpicture}[
  baseline,
  label distance=10pt % added
]
\matrix (mymatrix) [matrix of math nodes,left delimiter=(,right delimiter=),row sep=0.1cm,column sep=0.1cm, column 5/.style={red},  column 6/.style={red},  column 7/.style={red},  column 8/.style={red}, row 1/.style={nodes={font=\footnotesize}}, row 2/.style={nodes={font=\footnotesize}}, row 3/.style={nodes={font=\footnotesize}}, row 4/.style={nodes={font=\footnotesize}}, row 5/.style={nodes={font=\footnotesize}}] (m) {
0.50 & 0.50 & 0.50 & 0.50 & 0.50 & 0.50 & 0.50 & 0.50 \\ 
0.75 & 0.25 & 0.75 & 0.25 & 0.75 & 0.25 & 0.75 & 0.25 \\ 
0.25 & 0.75 & 0.25 & 0.75 & 0.25 & 0.75 & 0.25 & 0.75 \\ 
0.38 & 0.38 & 0.62 & 0.12 & 0.88 & 0.88 & 0.12 & 0.62 \\ 
0.88 & 0.88 & 0.12 & 0.62 & 0.38 & 0.38 & 0.62 & 0.12 \\ };

\node[
  fit=(m-1-1)(m-1-4),
  inner xsep=0,
  above delimiter=\{,
  label=above:$\bm{A}$
] {};

\node[
  fit=(m-1-5)(m-1-8),
  inner xsep=0,
  above delimiter=\{,
  label=above:$\bm{B}$
 ] {};
\end{tikzpicture} \\
\bm{A}_{B}^{(1)} & =
\begin{tikzpicture}[
  baseline,
  label distance=10pt % added
]
\matrix (mymatrix) [matrix of math nodes,left delimiter=(,right delimiter=),row sep=0.1cm,column sep=0.1cm,  column 1/.style={red}, row 1/.style={nodes={font=\footnotesize}}, row 2/.style={nodes={font=\footnotesize}}, row 3/.style={nodes={font=\footnotesize}}, row 4/.style={nodes={font=\footnotesize}}, row 5/.style={nodes={font=\footnotesize}}] (m) {
0.50 & 0.50 & 0.50 & 0.50 \\ 
0.75 & 0.25 & 0.75 & 0.25 \\ 
0.25 & 0.75 & 0.25 & 0.75 \\ 
0.88 & 0.38 & 0.62 & 0.12 \\ 
0.38 & 0.88 & 0.12 & 0.62 \\ };
\end{tikzpicture}  \\
\bm{A}_{B}^{(2)} & =
\begin{tikzpicture}[
  baseline,
  label distance=10pt % added
]
\matrix (mymatrix) [matrix of math nodes,left delimiter=(,right delimiter=),row sep=0.1cm,column sep=0.1cm,  column 2/.style={red}, row 1/.style={nodes={font=\footnotesize}}, row 2/.style={nodes={font=\footnotesize}}, row 3/.style={nodes={font=\footnotesize}}, row 4/.style={nodes={font=\footnotesize}}, row 5/.style={nodes={font=\footnotesize}}] (m) {
0.50 & 0.50 & 0.50 & 0.50 \\ 
0.75 & 0.25 & 0.75 & 0.25 \\ 
0.25 & 0.75 & 0.25 & 0.75 \\ 
0.38 & 0.88 & 0.62 & 0.12 \\ 
0.88 & 0.38 & 0.12 & 0.62 \\ };
\end{tikzpicture}  \\
& \vdots
\end{aligned} 
\label{eq:ab_matrices}
\end{equation}
\caption{Example of the creation of an $\bm{A}$, $\bm{B}$ and  $\bm{A}_{B}^{(i)}$ matrices. The $\bm{Q}$ matrix has been created with \cite{Sobol1967, Sobol1976} quasi-random numbers, $k=4$ and $N=5$. The figure is based on \cite{Puyj}.}
\label{fig:ab_matrices}
\end{figure}

In these matrices each column is a model input and each row a sampling
point. Any sampling point in either \(\bm{A}\) or \(\bm{B}\) can be
indicated as \(x_{vi}\), where \(v\) and \(i\) respectively index the
row (from 1 to \(N\)) and the column (from 1 to \(k\)).

First and total-order effects are then calculated by averaging several
elementary effects computed rowwise: for \(S_i\) we need pairs of points
where all factors but \(x_i\) have different values (i.e., \(\bm{A}_v\),
\((\bm{B}_A^{(i)})_v\); or \(\bm{B}_v\), \((\bm{A}_B^{(i)})_v\)), and
for \(T_i\) pairs of points where all factors except \(x_i\) have the
same values (i.e., \(\bm{A}_v\), \((\bm{A}_B^{(i)})_v\); or
\(\bm{B}_v\), \((\bm{B}_A^{(i)})_v\)). The elementary effect for \(S_i\)
thus requires moving from \(\bm{A}_v\) to \((\bm{B}_A^{(i)})_v\) (or
from \(\bm{B}_v\) to \((\bm{A}_B^{(i)})_v\)), therefore taking a step
along \(\bm{x}_{\sim i}\), whereby the elementary effect for \(T_i\)
involves moving from \(\bm{A}_v\) to \((\bm{A}_B^{(i)})_v\) (or from
\(\bm{B}_v\) to \((\bm{B}_A^{(i)})_v\)), hence moving along \(x_i\)
\citep{Saltelli2010a}. These pairs of points are the output \(y\)
obtained after running the model \(f\) in the \(v\)-th row of the
\(\bm{A}, \bm{B}\hdots\) matrices, denoted as
\(f(\bm{A})_v, f(\bm{B})_v,\hdots\).

The function \code{sobol_matrices()} allows to create these sampling
designs using either \cite{Sobol1967, Sobol1976} quasi-random numbers
(\code{type = "QRN"}), Latin hypercube sampling (\code{type = "LHS"}) or
random numbers (\code{type = "R"}). In Figure~\ref{fig:ab_matrices} we
show how these sampling methods differ in two dimensions. Comparatively,
quasi-random numbers fill the input space quicker and more evenly,
leaving smaller unexplored volumes. However, random numbers may provide
more accurate sensitivity indices when the model under examination has
important high-order terms \citep{Kucherenko2011}. Latin hypercube
sampling may outperform quasi-random numbers for some specific function
typologies. In general, quasi-random numbers are the safest bet when
selecting a sampling algorithm for a function of unknown behavior
\citep{Kucherenko2015b}, and are the default setting in \pkg{sensobol}.

\begin{CodeChunk}
\begin{figure}

{\centering \includegraphics{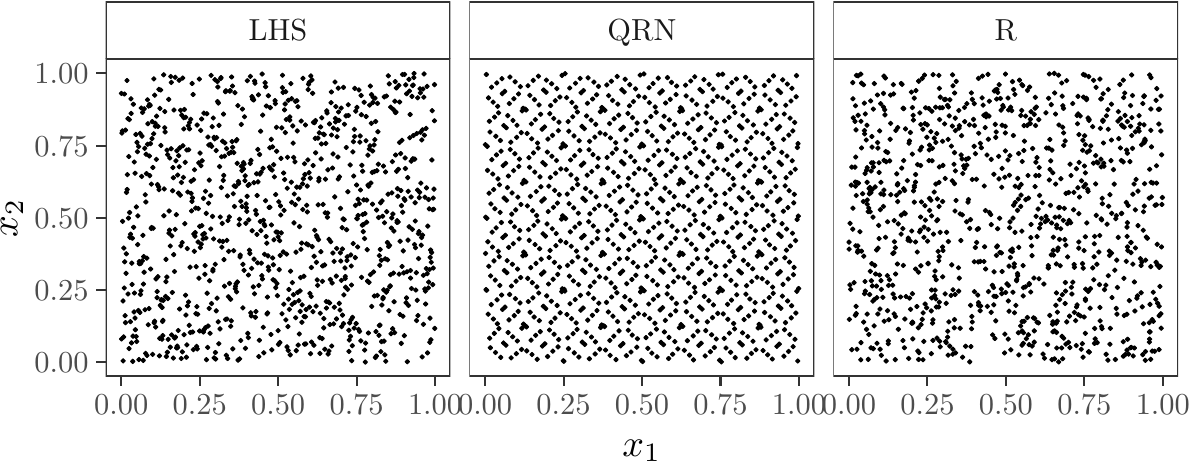} 

}

\caption[Sampling methods]{Sampling methods. Each dot is a sampling point. $N=2^{10}$.}\label{fig:visualization_matrices}
\end{figure}
\end{CodeChunk}

Once the sampling design is set, the computation of Sobol' indices is
done with the function \code{sobol_indices()}. The arguments
\code{first}, \code{total} and \code{matrices} are set by default at
\code{first = "saltelli"}, \code{total = "jansen"} and
\code{matrices = c("A", "B", "AB")} following best practices in
sensitivity analysis \citep{Saltelli2010a, Puyj}. However, any
combination between any of the first and total-order estimators listed in
Tables~\ref{tab:si_estimators}--\ref{tab:ti_estimators} is possible with
the appropriate sampling design (Table~\ref{tab:combinations}). If the
analyst selects estimators whose combination do not match the specific
designs listed in Table~\ref{tab:combinations}, \code{sobol_indices()}
will generate an error and urge to revise the specifications. This would
be the case, for instance, if the analyst sets \code{first = "sobol"},
\code{total = "glen"} and \code{matrices = "c("A", "AB", "BA")}.

\begingroup
\renewcommand{\arraystretch}{1.9}
\begin{table}[!ht]
\centering
\begin{tabular}{llll}
\toprule
\code{first} & \code{total} & \code{matrices} & Nº model runs \\
\midrule
\makecell{\code{"saltelli"} \\ \code{"jansen"}} & \makecell{\code{"jansen"}\\ \code{"sobol"} \\ \code{"homma"} \\ \code{"janon"} \\ \code{"glen"}} & \code{c("A", "B", "AB")} & $N(k + 2)$ \\
\midrule
\code{"sobol"} & \code{"saltelli"} & \code{c("A", "B", "BA")} & $N(k + 2)$ \\
\midrule
\code{"azzini"} & \makecell{\code{"jansen"} \\ \code{"sobol"} \\ \code{"homma"} \\ \code{"janon"} \\ \code{"glen"}\\ \code{"azzini"} \\ \code{"saltelli"}} & \code{c("A", "B", "AB", "BA")} & $2N(k + 1)$ \\
\midrule
\makecell{\code{"saltelli"} \\ \code{"jansen"}\\ \code{"sobol"} \\ \code{"azzini"}} & \code{"azzini"} & \code{c("A", "B", "AB", "BA")} & $2N(k + 1)$ \\
\bottomrule
\end{tabular}
\caption{\label{tab:combinations} Available combinations of first and total-order estimators in \pkg{sensobol} (v1.0.3).}
\end{table}
\endgroup

\section{Usage}
\label{sec:usage}

In this section we illustrate the functionality of \pkg{sensobol}
through three different examples of increasing complexity. Let us first
load the required packages:

\begin{CodeChunk}
\begin{CodeInput}
R> library("sensobol")
R> library("data.table")
R> library("ggplot2")
\end{CodeInput}
\end{CodeChunk}

\subsection{Example 1: The Sobol' G function}

In sensitivity analysis, the accuracy of sensitivity estimators is
usually checked against test functions for which the variance and the
sensitivity indices can be expressed analytically. \pkg{sensobol}
includes six of these test functions: \cite{Ishigami1990}'s,
\cite{Sobol1998}'s (known as G function), \cite{Bratley1992b}'s,
\cite{Bratley1988a}'s, \cite{Oakley2004}'s and \cite{Becker2020}'s
metafunction (Table~\ref{tab:test_functions}).

In this first example we illustrate the functionality of \pkg{sensobol}
with the Sobol' G function, one of the most used benchmark functions in
sensitivity analysis \citep{LoPiano2021, Puy2020, Saltelli2010a}. In its
current implementation, the Sobol' G is an eight-dimension function with
\(S_1>S_2>S_3>S_4\) and \((S_5,\hdots,S_8)\approx0\) (Table~\ref{tab:test_functions}, Nº 2). With this parametrization the Sobol' G
function is a Type A function according to \cite{Kucherenko2011}'s
taxonomy (a function with few important factors and minor interactions),
with Type B and Type C functions designing those with equally important
parameters but with few and large interactions respectively.

We first define the settings of the uncertainty and sensitivity
analysis: we set the sample size \(N\) of the base sample matrix and the
number of uncertain parameters \(k\), and create a vector with the
parameters' name. Since we will bootstrap the indices to get confidence
intervals, we set the number of bootstrap replicas to \(10^3\), the
bootstrap confidence interval method to the normal method and the
confidence intervals to 0.95:

\begingroup
\renewcommand{\arraystretch}{1.9}
\begin{table}[!ht]
\centering
\begin{tabular}{llp{3.2cm}}
\toprule
Nº & Test function & Author \\
\midrule
1 & \makecell{$y=\sin(x_1) +a \sin(x_2) ^ 2 + b x_3 ^4 \sin(x_1)$, \\ where $a=2, b=1$ and $(x_1,x_2,x_3)\sim\mathcal{U}(-\pi, +\pi)$} & \cite{Ishigami1990} \\
2 & \makecell{$y=\prod_{i=1}^{k} \frac{|4 x_i - 2| + a_i}{1 + a_i}$, \\ where $k=8$, $x_i\sim\mathcal{U}(0,1)$ and $a=(0, 1, 4.5, 9, 99, 99, 99, 99)$} & \cite{Sobol1998} \\
3 & \makecell{$y=\sum_{i=1}^{k}(-1)^i\prod_{j=1}^{i}x_j$, \\ where $x_i\sim\mathcal{U}(0,1)$} & \cite{Bratley1992b} \\
4 & \makecell{$y=\prod_{i=1}^{k} |4x_i - 2 |$, \\ where $x_i\sim\mathcal{U}(0,1)$} & \cite{Bratley1988a} \\
5 & \makecell{$y=\bm{a}_1^T \bm{x} + \bm{a}_2 ^ T sin(\bm{x}) + \bm{a}_3 ^ T cos(\bm{x}) + \bm{x}^T \bm{M}\bm{x}$, \\ where $\bm{x}=x_1,x_2,\hdots,x_k$, $k=15$, and values \\ for $\bm{a}^T_i,i=1,2,3$ and $\bm{M}$ are defined by the authors} & \cite{Oakley2004} \\
6 & $\begin{aligned}
y = & \sum_{i=1}^{k}\alpha_i f^{u_i}(x_i) + \sum_{i=1}^{k_2}\beta_i f^{u_{V_{i,1}}}(x_{V_{i,1}}) f^{u_{V_{i,2}}} (x_{V_{i,2}}) \\
& + \sum_{i=1}^{k_3}\gamma_i f^{u_{W_{i,1}}}(x_{W_{i,1}}) f^{u_{W_{i,2}}}(x_{W_{i,2}}) f^{u_{W_{i,3}}} (x_{W_{i,3}})
\end{aligned}$ & See \cite{Becker2020} and \cite{Puyj} for details. \\
\bottomrule
\end{tabular}
\caption{\label{tab:test_functions} Test functions included in \pkg{sensobol} (v1.0.3).}
\end{table}
\endgroup

\begin{CodeChunk}
\begin{CodeInput}
R> N <- 2 ^ 10
R> k <- 8
R> params <- paste("$x_", 1:k, "$", sep = "")
R> R <- 10^3
R> type <- "norm"
R> conf <- 0.95
\end{CodeInput}
\end{CodeChunk}

The next step is to create the sample matrix. In this specific case we
will use an \(\bm{A}\), \(\bm{B}\), \(\bm{A}_B^{(i)}\) design and Sobol'
quasi-random numbers to compute first and total-order indices. These are
default settings in \code{sobol_matrices()}. In our call to the function
we only need to define the sample size and the parameters:

\begin{CodeChunk}
\begin{CodeInput}
R> mat <- sobol_matrices(N = N, params = params)
\end{CodeInput}
\end{CodeChunk}

Once the sample matrix is defined we can run our model. Note that in
\code{mat} each column is a model input and each row a sample point,
hence the model has to be coded as to run rowwise. This is already the
case of the Sobol' G function included in \pkg{sensobol}:

\begin{CodeChunk}
\begin{CodeInput}
R> y <- sobol_Fun(mat)
\end{CodeInput}
\end{CodeChunk}

The package also allows the user to swiftly visualize the model output
uncertainty by plotting an histogram of the model output obtained from
the \(\bm{A}\) matrix:

\begin{CodeChunk}
\begin{CodeInput}
R> plot_uncertainty(Y = y, N = N) + labs(y = "Counts", x = "$y$")
\end{CodeInput}
\begin{figure}

{\centering \includegraphics{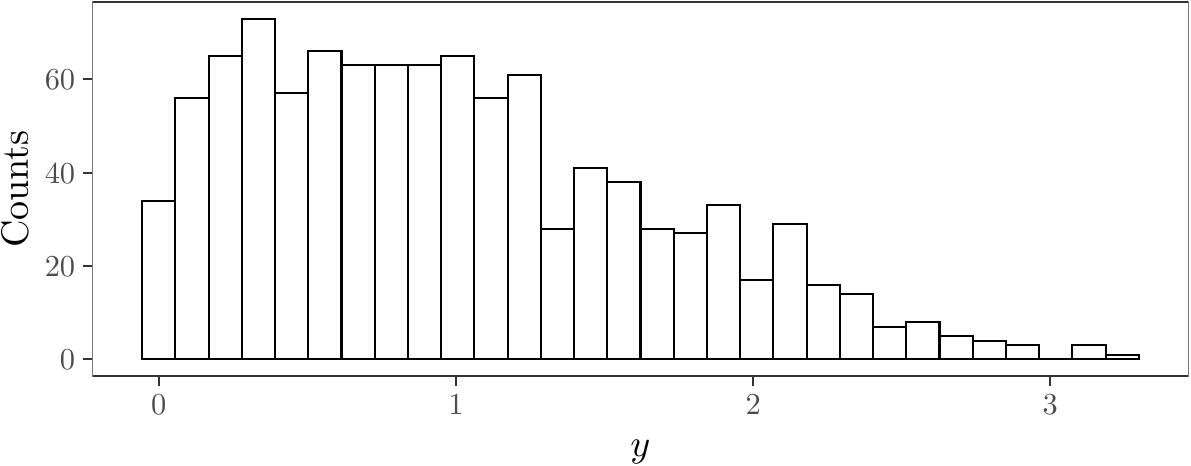} 

}

\caption[Empirical distribution of the Sobol' G model output]{Empirical distribution of the Sobol' G model output.}\label{fig:unc_sobolg}
\end{figure}
\end{CodeChunk}

Before computing Sobol' indices we recommend to explore how the model
output maps onto the model input space. \pkg{sensobol} includes two
functions to that aim, \code{plot_scatter()} and
\code{plot_multiscatter()}. The first displays the model output \(y\)
against \(x_i\) while showing the mean \(y\) value (i.e., as in Figures~\ref{fig:binned_mean}--\ref{fig:ishi_plot}), and allows the user to identify patterns denoting sensitivity
\citep{Pianosi2016}.

\begin{CodeChunk}
\begin{CodeInput}
R> plot_scatter(data = mat, N = N, Y = y, params = params)
\end{CodeInput}
\begin{figure}

{\centering \includegraphics{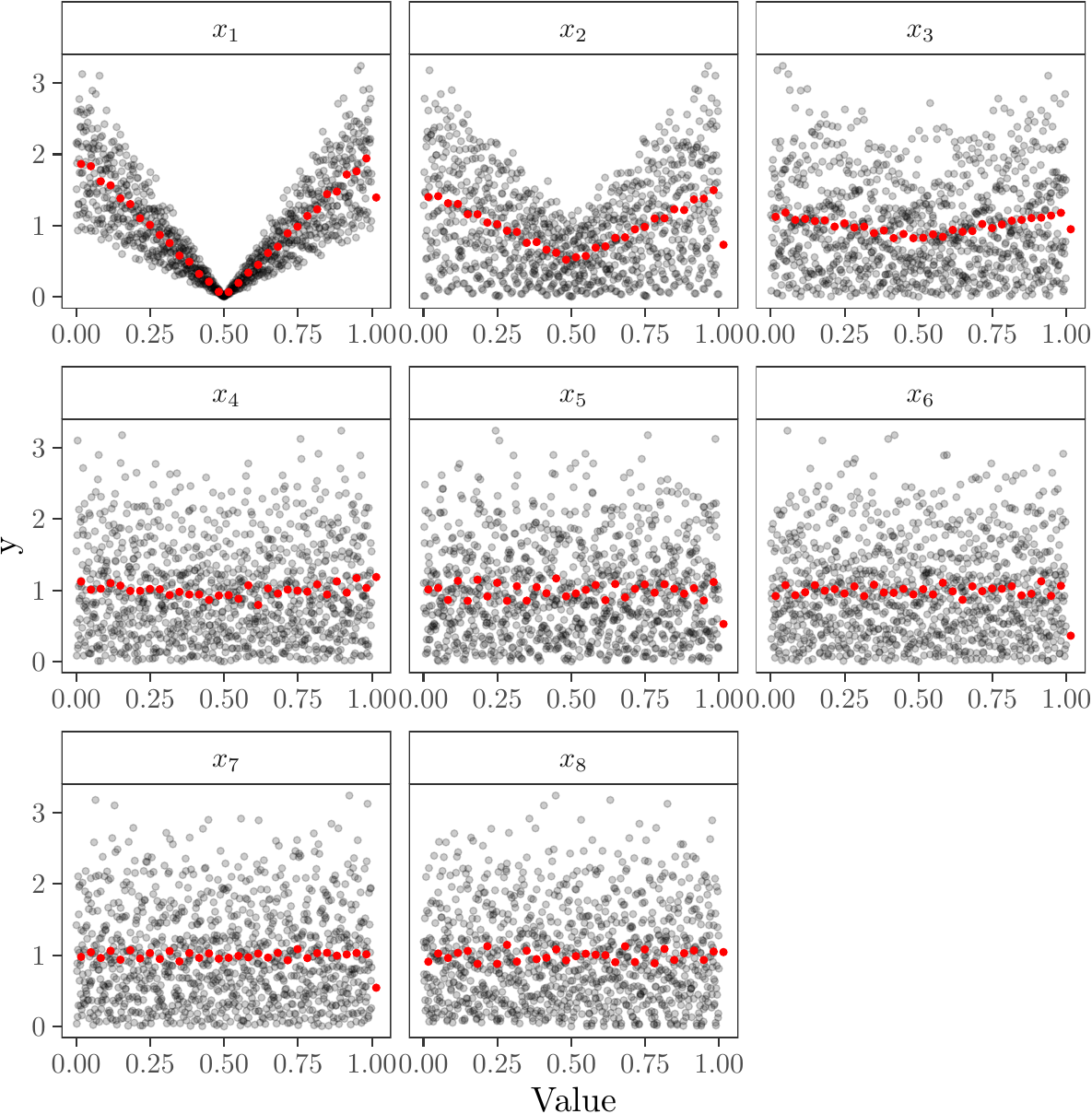} 

}

\caption[Scatter plots of model inputs against the model output for the Sobol' G function]{Scatter plots of model inputs against the model output for the Sobol' G function.}\label{fig:scatter_sobolg}
\end{figure}
\end{CodeChunk}

The scatter plots in Figure~\ref{fig:scatter_sobolg} evidence that \(x_1\), \(x_2\) and \(x_3\)
have more ``shape'' than the rest and thus have a higher influence on
\(y\) than \((x_4,\hdots,x_8)\). However, scatter plots do not always
permit to detect which parameters have a joint effect on the model
output. To gain a first insight on these interactions, the function
\code{plot_multiscatter()} plots \(x_i\) against \(x_j\) and maps the
resulting coordinate to its respective model output value. Interactions
are then visible by the emergence of colored patterns.

By default, \code{plot_multiscatter()} plots all possible combinations
of \(x_i\) and \(x_j\), which equal \(\frac{k!}{2!(k-2)!}=6\) possible
combinations in this specific case. In high-dimensional models with
several inputs this might lead to overplotting. To avoid this drawback,
the user can subset the parameters they wish to focus on following the
results obtained with \code{plot_scatter()}: if \(x_i\) does not show
``shape'' in the scatterplots of \(x_i\) against \(y\), then it may be
excluded from \code{plot_multiscatter()}.

Below we plot all possible combinations of pairs of inputs between
\(x_1-x_4\), which are influential according to Figure~\ref{fig:scatter_sobolg}

\begin{CodeChunk}
\begin{CodeInput}
R> plot_multiscatter(data = mat, N = N, Y = y, params = paste("$x_", 1:4, "$", 
+   sep = ""))
\end{CodeInput}
\begin{figure}

{\centering \includegraphics{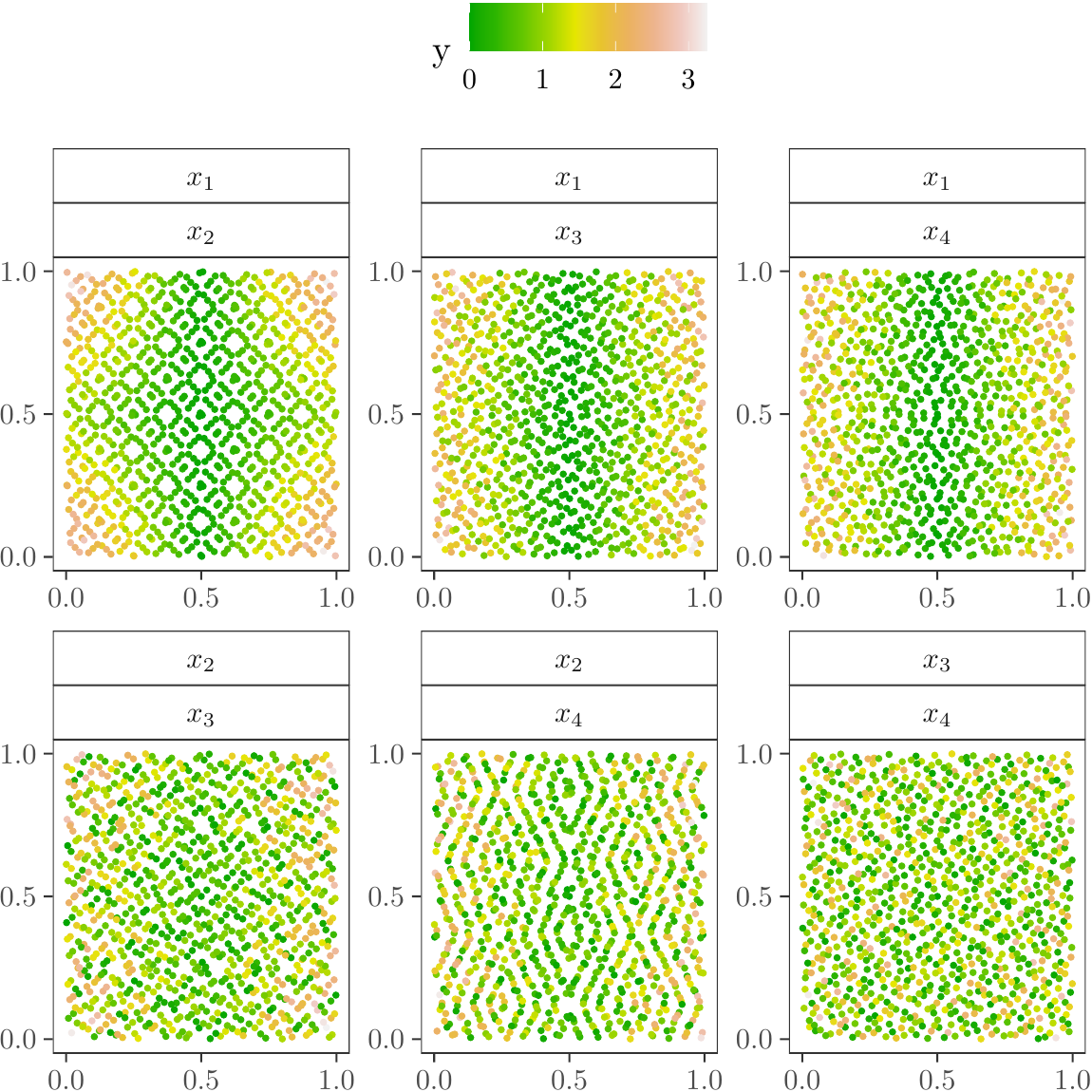} 

}

\caption[Scatter plot matrix of pairs of model inputs for the Sobol' G function]{Scatter plot matrix of pairs of model inputs for the Sobol' G function. The topmost and bottommost label facets refer to the $x$ and the $y$ axis respectively.}\label{fig:multiscatter_sobolg}
\end{figure}
\end{CodeChunk}

The results suggest that \(x_1\) might interact with \(x_2\) given the
colored pattern of the \((x_1, x_2)\) facet: the highest values of the
model output are concentrated in the corners of the (\(x_1,x_2\)) input
space and thus result from combinations of high/low \(x_1\) values with
high/low \(x_2\) values. In case the analyst is interested in assessing
the exact weight of this high-order interaction, the computation of
second-order indices would be required.

The last step is the computation of Sobol' indices. We set
\code{boot = TRUE} to bootstrap the Sobol' indices and get confidence
intervals:

\begin{CodeChunk}
\begin{CodeInput}
R> ind <- sobol_indices(Y = y, N = N, params = params, boot = TRUE, R = R, 
+   type = type, conf = conf)
\end{CodeInput}
\end{CodeChunk}

The output of \code{sobol_indices()} is an S3 object of class
\code{sensobol} with the results stored in the component \code{results}.
To improve the visualization of the object, we set the number of digits
in each numerical column to 3:

\begin{CodeChunk}
\begin{CodeInput}
R> cols <- colnames(ind$results)[1:5]
R> ind$results[, (cols):= round(.SD, 3), .SDcols = (cols)]
R> ind
\end{CodeInput}
\begin{CodeOutput}

First-order estimator: saltelli | Total-order estimator: jansen 

Total number of model runs: 10240 

Sum of first order indices: 0.9419303 
    original   bias std.error low.ci high.ci sensitivity parameters
 1:    0.724  0.005     0.069  0.584   0.854          Si      $x_1$
 2:    0.184 -0.002     0.039  0.110   0.261          Si      $x_2$
 3:    0.025  0.000     0.015 -0.005   0.054          Si      $x_3$
 4:    0.010  0.000     0.008 -0.007   0.025          Si      $x_4$
 5:    0.000  0.000     0.001 -0.001   0.002          Si      $x_5$
 6:    0.000  0.000     0.001 -0.002   0.002          Si      $x_6$
 7:    0.000  0.000     0.001 -0.001   0.002          Si      $x_7$
 8:    0.000  0.000     0.001 -0.002   0.002          Si      $x_8$
 9:    0.799  0.001     0.036  0.728   0.868          Ti      $x_1$
10:    0.243  0.000     0.013  0.217   0.269          Ti      $x_2$
11:    0.035  0.000     0.002  0.030   0.039          Ti      $x_3$
12:    0.011  0.000     0.001  0.009   0.012          Ti      $x_4$
13:    0.000  0.000     0.000  0.000   0.000          Ti      $x_5$
14:    0.000  0.000     0.000  0.000   0.000          Ti      $x_6$
15:    0.000  0.000     0.000  0.000   0.000          Ti      $x_7$
16:    0.000  0.000     0.000  0.000   0.000          Ti      $x_8$
\end{CodeOutput}
\end{CodeChunk}

The output informs of the first and total-order estimators used in the
calculation, the total number of model runs and the sum of the
first-order indices. If \(( \sum_{i=1}^{k} S_i) < 1\), the model is
non-additive.

When \code{boot = TRUE}, the output of \code{sobol_indices()} displays
the bootstrap statistics in the five leftmost columns (the observed
statistic, the bias, the standard error and the low and high confidence
intervals), and two extra columns linking each statistic to a
sensitivity index (\code{sensitivity}) and a parameter
(\code{parameters}). If \code{boot = FALSE}, \code{sobol_indices()}
computes a point estimate of the indices and the output includes only
the columns \code{original}, \code{sensitivity} and \code{parameters}.

The results indicate that \(x_1\) conveys 72\% of the uncertainty in
\(y\), followed by \(x_2\) (18\%). \(x_3\) and \(x_4\) have a very minor
first-order effect, while the rest are non-influential. Note the
presence of non-additivities: \(T_1\) and \(T_2\) (0.79 and 0.24) are
respectively higher than \(S_1\) and \(S_2\) (0.72 and 0.18). As we have
seen in Figure~\ref{fig:multiscatter_sobolg}, \(x_1\) and \(x_2\) have a non-additive effect on
\(y\).

We can also compute the Sobol' indices of a dummy parameter, i.e., a
parameter that has no influence on the model output, to estimate the
numerical approximation error. This will be used later on to identify
parameters whose contribution to the output variance is less than the
approximation error and hence can not be considered influential. Like
\code{sobol_indices()}, the function \code{sobol_dummy()} allows to
obtain point estimates (the default) or bootstrap estimates. In this
example we use the latter option:

\begin{CodeChunk}
\begin{CodeInput}
R> ind.dummy <- sobol_dummy(Y = y, N = N, params = params, boot = TRUE, 
+   R = R)
\end{CodeInput}
\end{CodeChunk}

The last stage is to plot the Sobol' indices and their confidence
intervals, as well as the Sobol' indices of a dummy parameter, with a
simple call to \code{plot}:

\begin{CodeChunk}
\begin{CodeInput}
R> plot(ind, dummy = ind.dummy)
\end{CodeInput}
\begin{figure}

{\centering \includegraphics{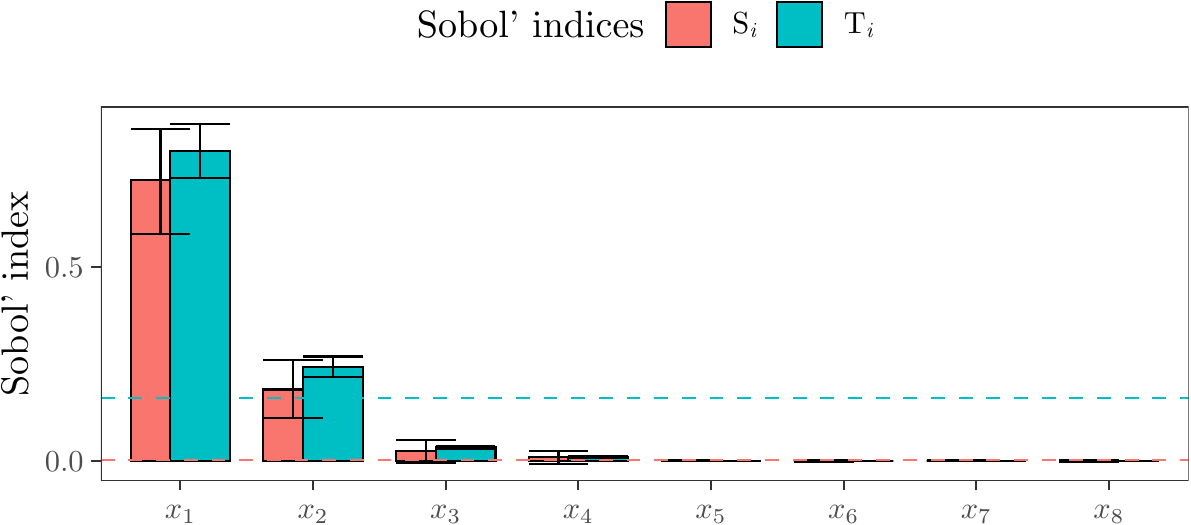} 

}

\caption[Sobol' indices of the Sobol' G function]{Sobol' indices of the Sobol' G function.}\label{fig:plot_indices_sobolg}
\end{figure}
\end{CodeChunk}

The error bars of \(S_1\) and \(S_2\) overlap with those of \(T_1\) and
\(T_2\) respectively. In the case of the Sobol' G function we know that
\(T_1>S_1\) and \(T_2>S_2\) because the analytic variance is known, but
for models where this is not the case such overlap might hamper the
identification of non-additivities. Narrower confidence intervals can be
obtained by increasing the sample size \(N\) and re-running the analysis
from the creation of the sample matrix onwards.

The horizontal, blue / red dashed lines respectively mark the upper
limit of the \(T_i\) and \(S_i\) indices of the dummy parameter. This
helps in identifying which parameters condition the model output given
the sample size constraints of the analysis. Only parameters whose lower
confidence intervals are not below the \(S_i\) and \(T_i\) indices of
the dummy parameter can be considered truly influential, in this case
\(x_1\) and \(x_2\). Note that although \(T_3\ne0\), the \(T_i\) index
of the dummy parameter is higher than \(T_3\) and therefore \(T_3\) can
not be distinguished from the approximation error.

\subsection{Example 2: A logistic population growth model}

In this section we show how \pkg{sensobol} can be implemented to conduct
a global uncertainty and sensitivity analysis of a dynamic model. To
illustrate the effect of high-order interactions and show
\pkg{sensobol}'s capacity to appraise second-order effects, we use the
discrete form of the classic logistic population growth model:

\begin{equation}
N_{t+1} = rN_t \left ( 1 - \frac{N_t}{K} \right )
\label{eq:population_growth}
\end{equation}

Malthusian models of population growth (i.e., exponential growth) can
not forever describe the growth of a population because resources are
limited and competitive pressures ultimately impose limits on growth.
Most ways to incorporate that limit to growth in models share similar
dynamics, and the most intuitive and widely used is the form proposed by
Verhulst in Equation~\ref{eq:population_growth}, which was popularized
in Ecology by \cite{Pearl1920}. In this model, the population \(N\) at
time \(t\) is dependent on the growth rate \(r\), the number of
individuals \(N\) and the carrying capacity \(K\), defined as the
maximum number of individuals that a given environment can sustain. When
\(N\) approaches \(K\), the population growth slows down until the
number of individuals converges to a constant (Figure~\ref{fig:dynamics_population}).

\begin{CodeChunk}
\begin{figure}

{\centering \includegraphics{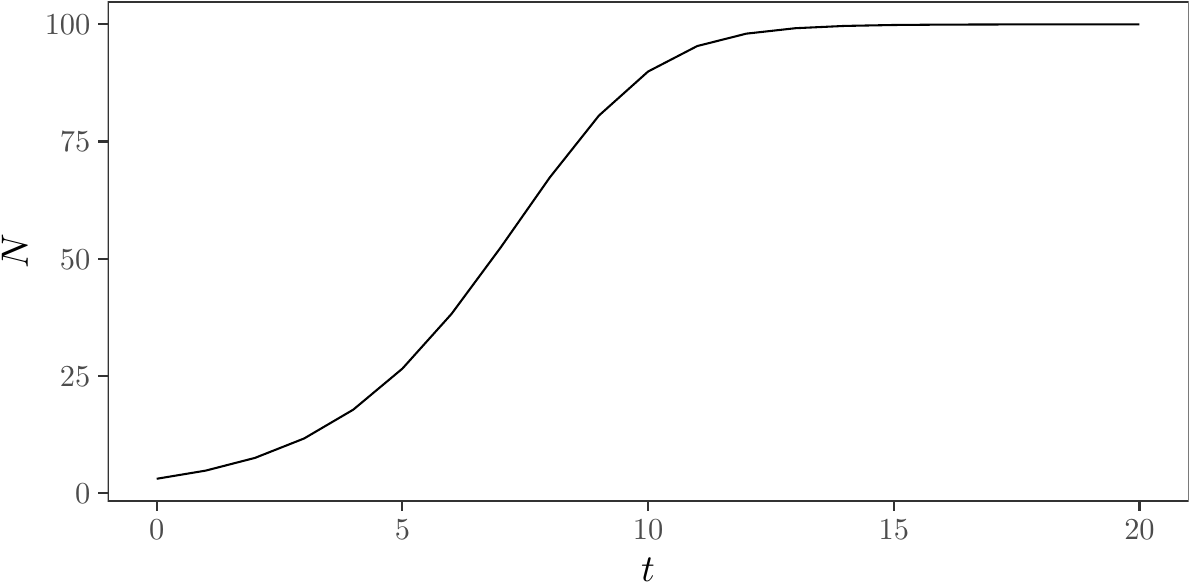} 

}

\caption[Dynamics of the logistic population growth model for $N_0=3$, $r = 0.6$ and $K=100$]{Dynamics of the logistic population growth model for $N_0=3$, $r = 0.6$ and $K=100$.}\label{fig:dynamics_population}
\end{figure}
\end{CodeChunk}

We first set the sample size \(N\) of the base sample matrix at
\(2^{13}\) and create a vector with the name of the parameters. For this
specific example we will use the \cite{Azzini2020} estimators, which
require a sampling design based on \(\bm{A}\), \(\bm{B}\),
\(\bm{A}_B^{(i)}\), \(\bm{B}_A^{(i)}\) matrices. We will compute up to
second-order effects, bootstrap the indices \(10^3\) times and compute
the 95\% confidence intervals using the percentile method.

\begin{CodeChunk}
\begin{CodeInput}
R> N <- 2 ^ 13
R> params <- c("$r$", "$K$", "$N_0$")
R> matrices <- c("A", "B", "AB", "BA")
R> first <- total <- "azzini"
R> order <- "second"
R> R <- 10 ^ 3
R> type <- "percent"
R> conf <- 0.95
\end{CodeInput}
\end{CodeChunk}

In the next two code snippets we code Equation~\ref{eq:population_growth} and wrap it up in a \code{mapply()} call to
make it run rowwise:

\begin{CodeChunk}
\begin{CodeInput}
R> population_growth <- function (r, K, X0) {
+   X <- X0
+   for (i in 0:20) {
+     X <- X + r * X * (1 - X / K)
+     }
+   return (X)
+ }
\end{CodeInput}
\end{CodeChunk}

\begin{CodeChunk}
\begin{CodeInput}
R> population_growth_run <- function (dt) {
+   return(mapply(population_growth, dt[, 1], dt[, 2], dt[, 3]))
+ }
\end{CodeInput}
\end{CodeChunk}

We now construct the sample matrix. In this example we set
\code{type = "LHS"} to use a Latin hypercube sampling design:

\begin{CodeChunk}
\begin{CodeInput}
R> mat <- sobol_matrices(matrices = matrices, N = N, params = params, 
+   order = order, type = "LHS")
\end{CodeInput}
\end{CodeChunk}

Let's assume that, after surveying the literature and conducting
fieldwork, we have agreed that the uncertainty in the model inputs can
be fairly approximated with the distributions presented in Table~\ref{tab:parameters_population}. Note that the use of a uniform
distribution assumes the existence of physical bounds for \(N_0\).
Distributions such as the log-normal may be more appropriate if the
interval is assumed to be less strict and the probability of occurrence
of some values is higher than others, yet they can produce outliers
prone to seriously bias the sensitivity analysis under small sample
sizes. Modelers often resort to uniform distributions when the quality
of knowledge available does not allow to make any judgement of that
sort. Ultimately, the selection of the distributions relies on the
authors' expertise and should be fully justified.

\begin{table*}[ht]
\centering
\begin{tabular}{llc}
\toprule
Parameter & Description & Distribution \\
\midrule
$r$ & Population growth rate & $\mathcal{N}(1.7, 0.3)$\\
$K$ & Maximum carrying capacity & $\mathcal{N}(40, 1)$\\
$N_0$ & Initial population size & $\mathcal{U}(10, 50)$\\
\bottomrule
\end{tabular}
\caption{Summary of the parameters and their distributions \citep{Chalom2017}.}
\label{tab:parameters_population}
\end{table*}

We transform each model input in \code{mat} to its specific probability
distribution:

\begin{CodeChunk}
\begin{CodeInput}
R> mat[, "$r$"] <- qnorm(mat[, "$r$"], 1.7, 0.3)
R> mat[, "$K$"] <- qnorm(mat[, "$K$"], 40, 1)
R> mat[, "$N_0$"] <- qunif(mat[, "$N_0$"], 10, 50)
\end{CodeInput}
\end{CodeChunk}

The sample matrix in \code{mat} is now ready and we can run the model:

\begin{CodeChunk}
\begin{CodeInput}
R> y <- population_growth_run(mat)
\end{CodeInput}
\end{CodeChunk}

And display the model output uncertainty:

\begin{CodeChunk}
\begin{CodeInput}
R> plot_uncertainty(Y = y, N = N) + labs(y = "Counts", x = "$y$")
\end{CodeInput}
\begin{figure}

{\centering \includegraphics{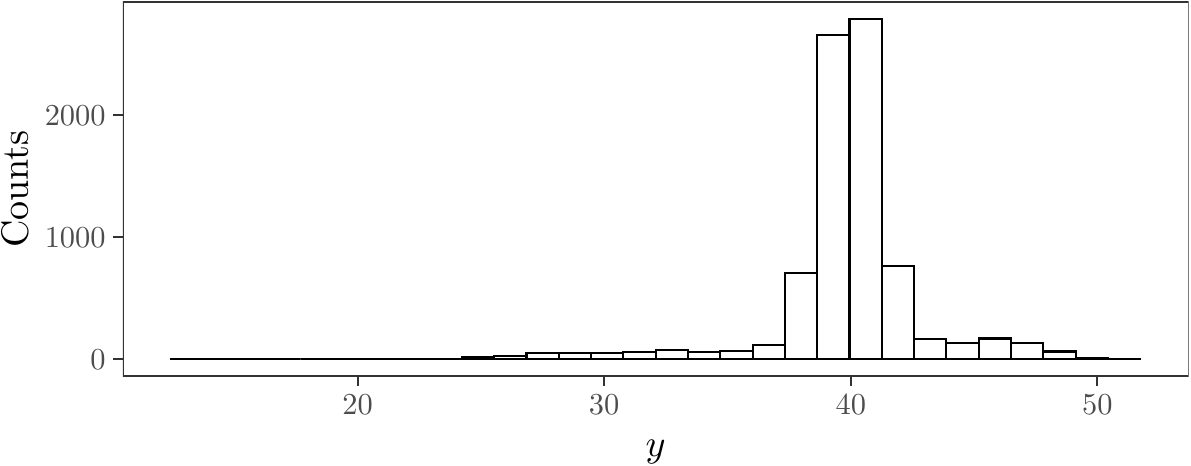} 

}

\caption[Empirical distribution of the logistic population growth model output]{Empirical distribution of the logistic population growth model output.}\label{fig:unc_population}
\end{figure}
\end{CodeChunk}

After 20 time steps the number of individuals will most likely
concentrate around 40. Note however the right and left tails of the
distribution, indicating that a few simulations also yielded
significantly lower and higher population values. These tails result
from some specific combinations of parameter values and are indicative
of interaction effects, which will be explored later on.

We can also compute some statistics to get a better grasp of the output
distribution, such as quantiles:

\begin{CodeChunk}
\begin{CodeInput}
R> quantile(y, probs = c(0.01, 0.025, 0.5, 0.975, 0.99, 1))
\end{CodeInput}
\begin{CodeOutput}
      1%     2.5%      50%    97.5%      99%     100% 
27.80714 30.66101 40.00111 46.64511 47.91589 53.41604 
\end{CodeOutput}
\end{CodeChunk}

With \code{plot_scatter()} we can map the model inputs onto the model
output. Instead of plotting one dot per simulation, in this example we
use hexagon bins by setting \code{method = "bin"} and internally calling
\code{ggplot2::geom_hex()}. With this specification we divide the plane
into regular hexagons, count the number of hexagons and map the number
of simulations to the hexagon bin. \code{method = "bin"} is a useful
resource to avoid overplotting with \code{plot_scatter()} when the
sample size of the base sample matrix (\(N\)) is high, as in this case
(\(N=2^{13}\)):

\begin{CodeChunk}
\begin{CodeInput}
R> plot_scatter(data = mat, N = N, Y = y, params = params, method = "bin")
\end{CodeInput}
\begin{figure}

{\centering \includegraphics{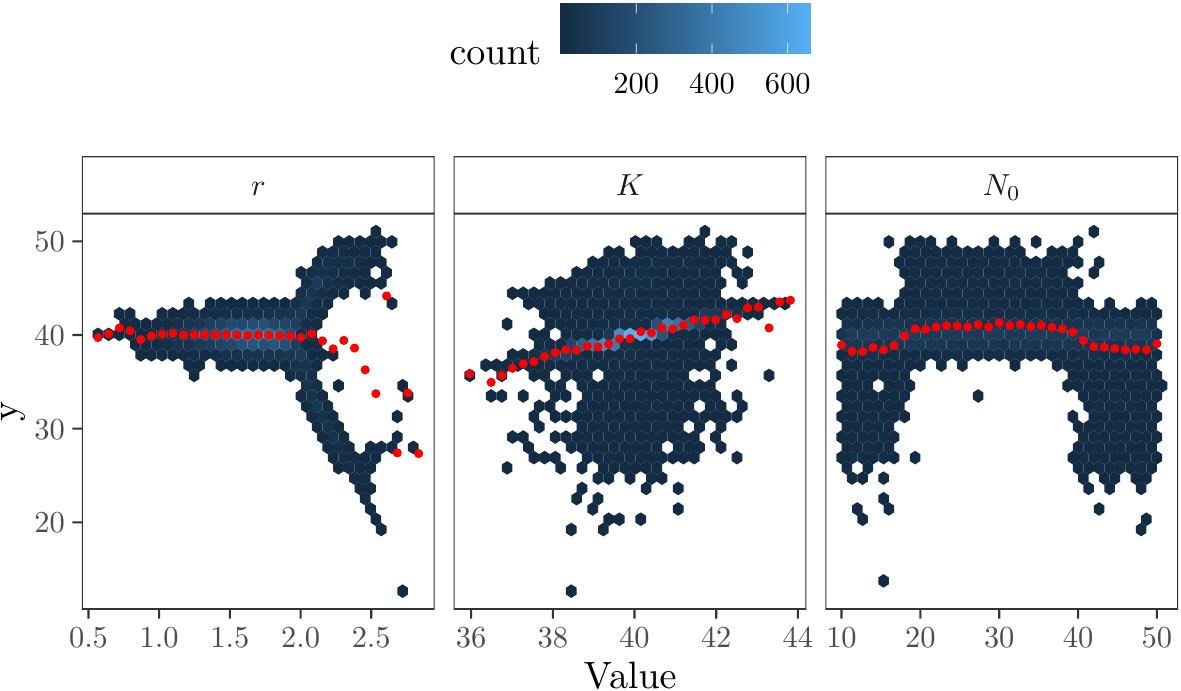} 

}

\caption[Hexbin plot of model inputs against the model output for the population growth model]{Hexbin plot of model inputs against the model output for the population growth model.}\label{fig:scatter_population}
\end{figure}
\end{CodeChunk}

Zones with a higher density of dots are highlighted by lighter blue
colors. Note also the bifurcation of the model output at \(r\approx2\).
This behavior is the result of the discretization of the logistic
\citep{May1976}. At \(r>2\), a cycle of length 2 emerges, followed as
\(r\) is increased further by an infinite sequence of period-doubling
bifurcations approaching a critical value after which chaotic behavior
and strange attractors result.

We can also avoid overplotting in \code{plot_multiscatter()} by randomly
sampling and displaying only \(n\) simulations. This is controlled by
the argument \code{smpl}, which is \code{NULL} by default. Here we set
\code{smpl} at \(2^{11}\) and plot only 1/4th of the simulations.

\begin{CodeChunk}
\begin{CodeInput}
R> plot_multiscatter(data = mat, N = N, Y = y, params = params, smpl = 2^11)
\end{CodeInput}
\begin{figure}

{\centering \includegraphics{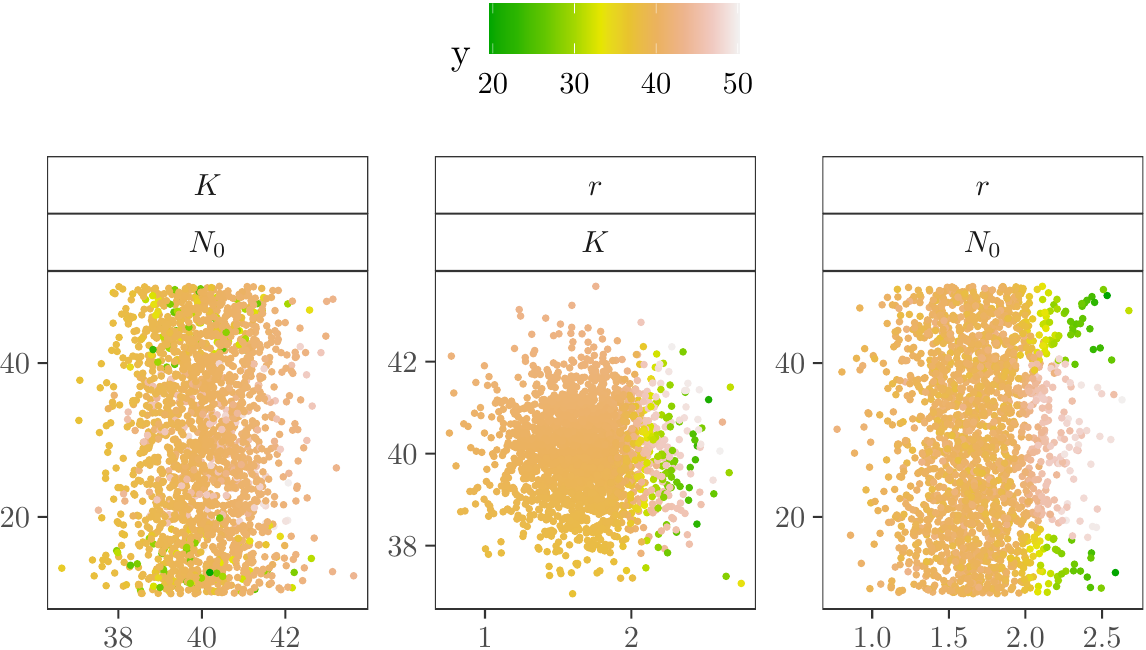} 

}

\caption[Scatterplot matrix of pairs of model inputs for the population growth model]{Scatterplot matrix of pairs of model inputs for the population growth model.}\label{fig:multiscatter_population}
\end{figure}
\end{CodeChunk}

The results suggest that there may be interactions between \((r,K)\) and
\((r,N_0)\): note the yellow-green colours concentrated on the right
side of the \((r,K)\) plot as well as on the top right and bottom right
sides of the \((r,N_0)\) plot.

These interactions, as well as the first-order effects of \(N_0,r,K\),
can be quantified with a call to \code{sobol_indices()}:

\begin{CodeChunk}
\begin{CodeInput}
R> ind <- sobol_indices(matrices = matrices, Y = y, N = N, params = params, 
+   first = first, total = total, order = order, boot = TRUE, R = R, 
+   parallel = "no", type = type, conf = conf)
\end{CodeInput}
\end{CodeChunk}

We round the number of digits of the numeric columns to 3 to better
inspect the results:

\begin{CodeChunk}
\begin{CodeInput}
R> cols <- colnames(ind$results)[1:5]
R> ind$results[, (cols):= round(.SD, 3), .SDcols = (cols)]
R> ind
\end{CodeInput}
\begin{CodeOutput}

First-order estimator: azzini | Total-order estimator: azzini 

Total number of model runs: 114688 

Sum of first order indices: 0.2569059 
   original   bias std.error low.ci high.ci sensitivity parameters
1:    0.028 -0.001     0.019 -0.010   0.067          Si        $r$
2:    0.114  0.000     0.004  0.107   0.122          Si        $K$
3:    0.115  0.000     0.011  0.093   0.136          Si      $N_0$
4:    0.760  0.000     0.011  0.737   0.781          Ti        $r$
5:    0.185  0.000     0.010  0.164   0.206          Ti        $K$
6:    0.861  0.001     0.020  0.825   0.900          Ti      $N_0$
7:   -0.004  0.000     0.010 -0.024   0.015         Sij    $r$.$K$
8:    0.673  0.001     0.022  0.627   0.718         Sij  $r$.$N_0$
9:    0.012  0.000     0.007 -0.002   0.025         Sij  $K$.$N_0$
\end{CodeOutput}
\end{CodeChunk}

The output also displays the second-order effects (\(S_{ij}\)) of the
pairs \((r,K)\), \((r,N_0)\) and \((K, N_0)\). Note that \(S_{r,N_0}\)
conveys \(\sim67\)\% of the uncertainty in \(y\), and that
\(S_r + S_K + S_{N_0}=2.8 + 11.4 + 11.5 \approx 25\)\%, meaning that
first-order effects are responsible for only \textit{circa} 1/4 th of
the model output variance. The model behaviour is largely driven by a
coupled effect which would have passed unnoticed should we had relied on
a local sensitivity analysis approach, i.e., if we had moved one
parameter at-a-time.

In any case, \(K\) and \(N_0\) have the higher first-order effect in the
model output. If the aim is to reduce the uncertainty in the prediction
(i.e., to better assess the potential impact of a species on a
territory), these results suggest to focus the efforts on better
quantifying the initial population \(N_0\), and/or on conducting further
research on what is the maximum carrying capacity \(K\) of the
environment for this particular species. Priority should be given to
\(N_0\) given its strong interaction with \(r\).

In order to get an estimation of the approximation error we compute the
Sobol' indices also for the dummy parameter:

\begin{CodeChunk}
\begin{CodeInput}
R> ind.dummy <- sobol_dummy(Y = y, N = N, params = params, boot = TRUE, 
+   R = R)
\end{CodeInput}
\end{CodeChunk}

And plot the output:

\begin{CodeChunk}
\begin{CodeInput}
R> plot(ind, dummy = ind.dummy)
\end{CodeInput}
\begin{figure}

{\centering \includegraphics{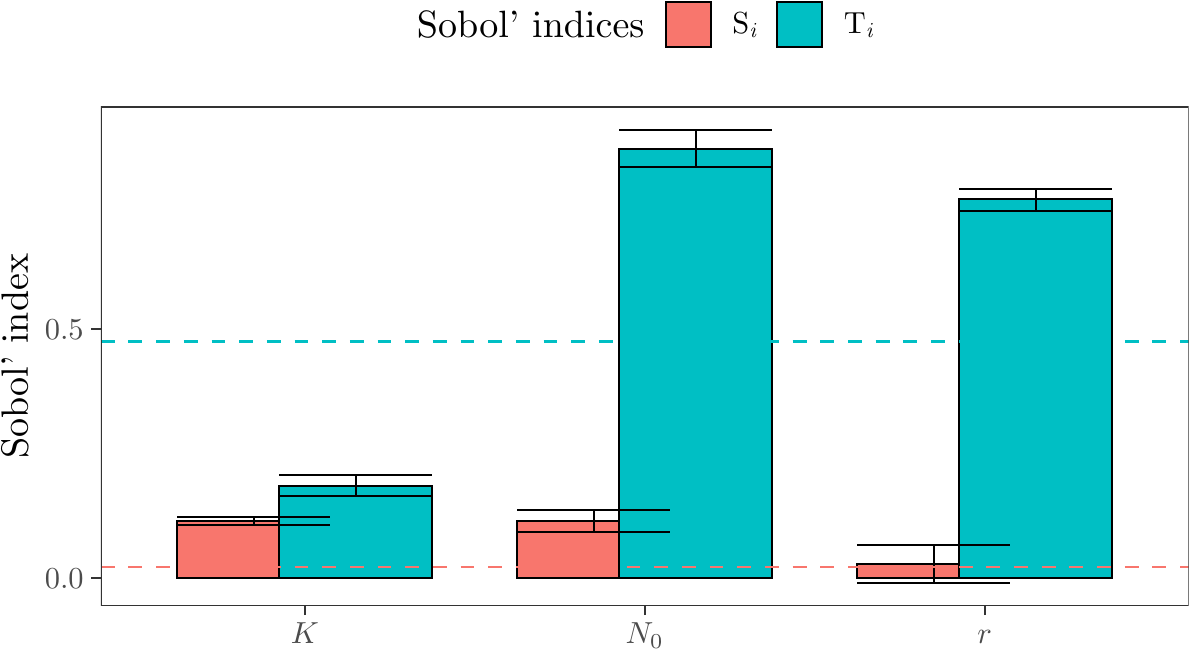} 

}

\caption[First and total-order Sobol' indices of the population growth model]{First and total-order Sobol' indices of the population growth model.}\label{fig:plot_indices_population}
\end{figure}
\end{CodeChunk}

Note the importance of interactions, reflected in \(T_{N_0}\gg S_{N_0}\)
and \(T_{r}\gg S_{r}\). It is also important to highlight that \(T_K\)
is below the \(T_i\) index of the dummy parameter (the dashed,
horizontal blue line at c.~0.5). This makes \(T_K\) indistinguishable
from the approximation error. The same applies to \(S_r\), whose 95\%
confidence interval overlaps with the \(S_i\) of the dummy parameter
(the dashed, red line).

Finally, we can also plot second-order indices by setting
\code{order = "second"} in a \code{plot()} call:

\begin{CodeChunk}
\begin{CodeInput}
R> plot(ind, order = "second")
\end{CodeInput}
\begin{figure}

{\centering \includegraphics{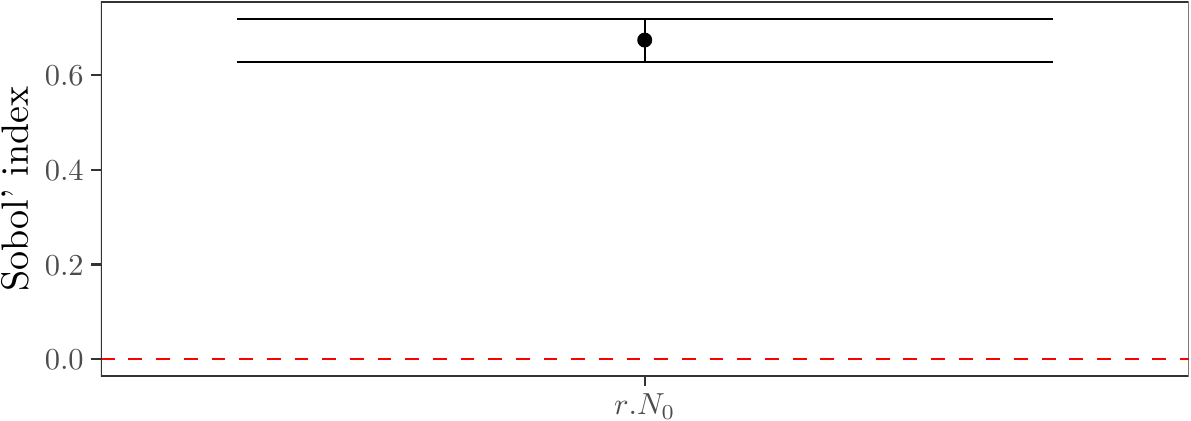} 

}

\caption[Second-order Sobol' indices]{Second-order Sobol' indices.}\label{fig:plot_indices2_population}
\end{figure}
\end{CodeChunk}

Only \(S_{r,N_0}\) is displayed because \code{plot()} only returns
second-order indices for which the lower confidence interval does not
overlap with zero.

\subsection{Example 3: The spruce budworm and forest model}
\label{sec:budworm}

This last example illustrates the flexibility of \pkg{sensobol} against
systems of differential equations. Under these circumstances, the
analyst might be interested in exploring the sensitivity of each model
output or state variable to the uncertain inputs at different points in
time, i.e., at the transient phase and/or at equilibrium. \pkg{sensobol}
integrates with the \pkg{deSolve} \citep{Soetaert2010} and the
\pkg{data.table} \citep{Dowle2020} packages to achieve this goal in an
easy manner.

We use the spruce budworm and forest model of \cite{Ludwig1976}. Spruce
budworm is a devastating pest of Canadian and high-latitude US balsam
fir and spruce forests. A half-century ago, research teams led by
Crawford Holling developed detailed models of the interaction between
the budworm and their target species, models capable of reproducing the
boom-and-bust dynamics and spatial patterns exhibited in real forests.
Donald Ludwig pointed out that these models were overparameterized and
that much simpler models could capture the essential dynamics in a more
robust manner.

The basic idea of the model is that the dynamics of the system play out
on multiple time scales. Budworm population dynamics respond to forest
quality on a fast time scale, leading to changes in forest quality on a
slower time scale. In turn, the slow dynamics change the topological
profile of the fast-time scale dynamics, introducing hysteretic
oscillations reminiscent of relaxation oscillations. The simplest
version of the budworm model in \cite{Ludwig1976} can be
non-dimensionalized easily, making transparent the reduction in
dimension on the fast-time scale. In place of those, however, we
consider the more complicated version given by Equations 20-22 in that
paper, yielding the explicit form

\begin{equation} 
\begin{split}
\frac{dB}{dt}&=r_B B\left(1 - \frac{B}{KS} \frac{T_E^2 + E^2}{E^2}\right) - \beta \frac{B^2}{(\alpha S)^2 + B^2} \\ 
\frac{dS}{dt} &= r_S S\left(1 - \frac{SK_E}{EK_S}\right) \\ 
\frac{dE}{dt} & = r_E E\left(1 - \frac{E}{K_E}\right) - P \frac{B}{S} \frac{E^2}{T_E^2 + E^2}
\end{split} \,,
\label{eq:budworm_model}
\end{equation}

where \(B\), \(S\) and \(E\) are the budworm density, the average size
of the trees and the energy reserve of the trees respectively (Figure~\ref{fig:plot_dynamics_budworm}). Equation~\ref{eq:budworm_model} allows a full characterization of the parameter space
with empirical data (Table~\ref{tab:parameters_budworm}).

\begin{table*}[ht]
\centering
\begin{tabular}{llc}
\toprule
Parameter & Description & Distribution \\
\midrule
$r_B$ & Intrinsic budworm growth rate & $\mathcal{U}(1.52, 1.6)$\\
$K$ & Maximum budworm density & $\mathcal{U}(100, 355)$\\
$\beta$ & Maximum budworm predated & $\mathcal{U}(20000, 43200)$\\
$\alpha$ & $\frac{1}{2}$ maximum density for predation & $\mathcal{U}(1, 2)$\\
$r_S$ & Intrinsic branch growth rate & $\mathcal{U}(0.095, 0.15)$\\
$K_S$ & Maximum branch density & $\mathcal{U}(24000, 25440)$\\
$K_E$ & Maximum $E$ level & $\mathcal{U}(1, 1.2)$\\
$r_E$ & Intrinsic $E$ growth rate & $\mathcal{U}(0.92, 1)$\\
$P$ & Consumption rate of $E$ & $\mathcal{U}(0.0015, 0.00195)$\\
$T_E$ & $E$ proportion & $\mathcal{U}(0.7, 0.9)$\\
\bottomrule
\end{tabular}
\caption{Summary of the parameters and their distribution in \cite{Ludwig1976}.}
\label{tab:parameters_budworm}
\end{table*}

Like in the previous examples, we first define the sample size of the
base sample matrix, a vector with the name of the parameters, the
order of the sensitivity indices investigated, the number of bootstrap
replicas and the type of confidence intervals. We plan to run the model
for 150 months at a 1 month interval (\code{times}) and extract the
model output every 25 months (\code{timeOutput}). Such settings have
been selected to get an insight into all the stages of the model (i.e.,
growth, equilibirum) (Figure~\ref{fig:plot_dynamics_budworm}).

\begin{CodeChunk}
\begin{CodeInput}
R> N <- 2 ^ 9
R> params <- c("r_b", "K", "beta", "alpha", "r_s", "K_s", "K_e", "r_e", "P", 
+   "T_e")
R> order <- "first"
R> R <- 10 ^ 3
R> type <- "norm"
R> conf <- 0.95
R> times <- seq(0, 150, 1)
R> timeOutput <- seq(25, 150, 25)
\end{CodeInput}
\end{CodeChunk}

Since the model in Equation~\ref{eq:budworm_model} is a system of
differential equations, we can code it following the guidelines set by
the \pkg{deSolve} package \citep{Soetaert2010}:

\begin{CodeChunk}
\begin{CodeInput}
R> budworm_fun <- function(t, state, parameters) {
+   with(as.list(c(state, parameters)), {
+     dB <- r_b * B * (1 - B / (K * S) * (T_e^2 + E^2) / E^2) - beta * 
+       B^2/((alpha ^ S)^2 + B^2)
+     dS <- r_s * S * (1 - (S * K_e) / (E * K_s))
+     dE <- r_e * E * (1 - E / K_e) - P * (B / S) * E^2/(T_e^2 + E^2)
+     list(c(dB, dS, dE))
+   })
+ }
\end{CodeInput}
\end{CodeChunk}

We can then create the sample matrix as in the previous examples:

\begin{CodeChunk}
\begin{CodeInput}
R> mat <- sobol_matrices(N = N, params = params, order = order)
\end{CodeInput}
\end{CodeChunk}

And transform each column to the probability distributions specified in
Table~\ref{tab:parameters_budworm}:

\begin{CodeChunk}
\begin{CodeInput}
R> mat[, "r_b"] <- qunif(mat[, "r_b"], 1.52, 1.6)
R> mat[, "K"] <- qunif(mat[, "K"], 100, 355)
R> mat[, "beta"] <- qunif(mat[, "beta"], 20000, 43200)
R> mat[, "alpha"] <- qunif(mat[, "alpha"], 1, 2)
R> mat[, "r_s"] <- qunif(mat[, "r_s"], 0.095, 0.15)
R> mat[, "K_s"] <- qunif(mat[, "K_s"], 24000, 25440)
R> mat[, "K_e"] <- qunif(mat[, "K_e"], 1, 1.2)
R> mat[, "r_e"] <- qunif(mat[, "r_e"], 0.92, 1)
R> mat[, "P"] <- qunif(mat[, "P"], 0.0015, 0.00195)
R> mat[, "T_e"] <- qunif(mat[, "T_e"], 0.7, 0.9)
\end{CodeInput}
\end{CodeChunk}

\begin{CodeChunk}
\begin{figure}

{\centering \includegraphics{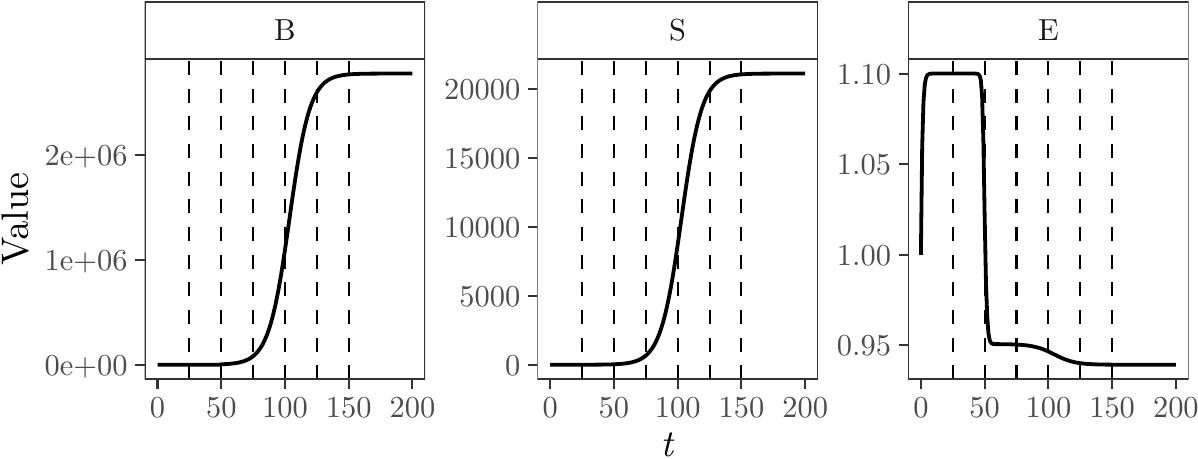} 

}

\caption[Dynamics of the spruce budworm and forest model]{Dynamics of the spruce budworm and forest model. The vertical, dashed lines mark the times at which we will conduct the sensitivity analysis. Initial state values: $B=1,S=0.07,E=1$. The parameter values are the mean values of the distributions shown in Table~\ref{tab:parameters_budworm}.}\label{fig:plot_dynamics_budworm}
\end{figure}
\end{CodeChunk}

We arrange a parallel setting to speed up the computations. To that aim,
we load the packages \code{foreach} \citep{Microsoft2020},
\code{parallel} \citep{RCoreTeam2020}, and \code{doParallel}
\citep{Microsoft2020a}.

\begin{CodeChunk}
\begin{CodeInput}
R> library("foreach")
R> library("parallel")
R> library("doParallel")
\end{CodeInput}
\end{CodeChunk}

In the next code snippet we design a loop to conduct the computations
rowwise. Note that the function \code{budworm_fun()} is called through
\pkg{sensobol}'s \code{sobol_ode()}, a wrapper around \pkg{deSolve}'s
\code{ode} function which allows to retrieve the model output at the
times specified in \code{timeOutput}. Before executing the nested loop
we order the computer to use 75\% of the cores available in order to
take advantage of parallel computing.

\begin{CodeChunk}
\begin{CodeInput}
R> n.cores <- makeCluster(floor(detectCores() * 0.75))
R> registerDoParallel(n.cores)
R> y <- foreach(i = 1:nrow(mat), .combine = "rbind", 
+   .packages = "sensobol") %dopar%
+   {
+   sobol_ode(d = mat[i, ], times = times, timeOutput = timeOutput,
+   state = c(B = 0.1, S = 007, E = 1), func = budworm_fun)
+   }
R> stopCluster(n.cores)
\end{CodeInput}
\end{CodeChunk}

Now we can rearrange the data for the sensitivity analysis. We first
convert the output to a \code{data.table} format:

\begin{CodeChunk}
\begin{CodeInput}
R> full.dt <- data.table(y)
R> print(full.dt)
\end{CodeInput}
\begin{CodeOutput}
       time          B          S         E
    1:   25   18257.52   148.6977 0.9505572
    2:   50  344848.35  2784.0488 0.9493628
    3:   75 2100795.05 16205.2496 0.9423155
    4:  100 2742807.83 20823.9167 0.9393914
    5:  125 2780918.41 21093.6378 0.9392101
   ---                                     
36860:   50  657250.10  6612.1482 1.0072944
36861:   75 2138101.78 20294.7121 0.9984543
36862:  100 2271563.34 21453.5366 0.9975345
36863:  125 2275187.60 21484.8468 0.9975091
36864:  150 2275280.24 21485.6470 0.9975085
\end{CodeOutput}
\end{CodeChunk}

And transform the resulting \code{data.table} from wide to long format:

\begin{CodeChunk}
\begin{CodeInput}
R> indices.dt <- melt(full.dt, measure.vars = c("B", "S", "E"))
R> print(indices.dt)
\end{CodeInput}
\begin{CodeOutput}
        time variable        value
     1:   25        B 1.825752e+04
     2:   50        B 3.448484e+05
     3:   75        B 2.100795e+06
     4:  100        B 2.742808e+06
     5:  125        B 2.780918e+06
    ---                           
110588:   50        E 1.007294e+00
110589:   75        E 9.984543e-01
110590:  100        E 9.975345e-01
110591:  125        E 9.975091e-01
110592:  150        E 9.975085e-01
\end{CodeOutput}
\end{CodeChunk}

With this transformation and the compatibility of \pkg{sensobol} with
the \pkg{data.table} package \citep{Dowle2020}, we can easily compute
variance-based sensitivity indices at each selected time step for each
state variable. We first activate 75\% of the CPUs to bootstrap the
Sobol' indices in parallel and then compute the Sobol' indices grouping
by \code{time} and \code{variable}:

\begin{CodeChunk}
\begin{CodeInput}
R> ncpus <- floor(detectCores() * 0.75)
R> indices <- indices.dt[, sobol_indices(Y = value, N = N, params = params,
+   order = order, boot = TRUE, first = "jansen", R = R,
+   parallel = "multicore", ncpus = ncpus)$results, .(variable, time)]
\end{CodeInput}
\end{CodeChunk}

We also compute the Sobol' indices of the dummy parameter:

\begin{CodeChunk}
\begin{CodeInput}
R> indices.dummy <- indices.dt[, sobol_dummy(Y = value, N = N, 
+   params = params), .(variable, time)]
\end{CodeInput}
\end{CodeChunk}

Finally, with some lines of code we can visualize the evolution of
\(S_i\) and \(T_i\) indices through time for each state variable and
uncertain model input:

\begin{CodeChunk}
\begin{CodeInput}
R> ggplot(indices, aes(time, original, fill = sensitivity,
+                     color = sensitivity,
+                     group = sensitivity)) +
+   geom_line() +
+   geom_ribbon(aes(ymin = indices[sensitivity %in% c("Si", "Ti")]$low.ci,
+                   ymax = indices[sensitivity %in% c("Si", "Ti")]$high.ci,
+                   color = sensitivity),
+               alpha = 0.1, linetype = 0) +
+   geom_hline(data = indices.dummy[, parameters:= NULL][sensitivity == "Ti"],
+              aes(yintercept = original, color = sensitivity, group = time),
+                       lty = 2, size = 0.1) +
+   guides(linetype = FALSE, color = FALSE) +
+   facet_grid(parameters ~ variable) +
+   scale_y_continuous(breaks = scales::pretty_breaks(n = 3)) +
+   labs(x = expression(italic(t)),
+        y = "Sobol' indices") +
+   theme_AP() +
+   theme(legend.position = "top")
\end{CodeInput}
\begin{figure}

{\centering \includegraphics{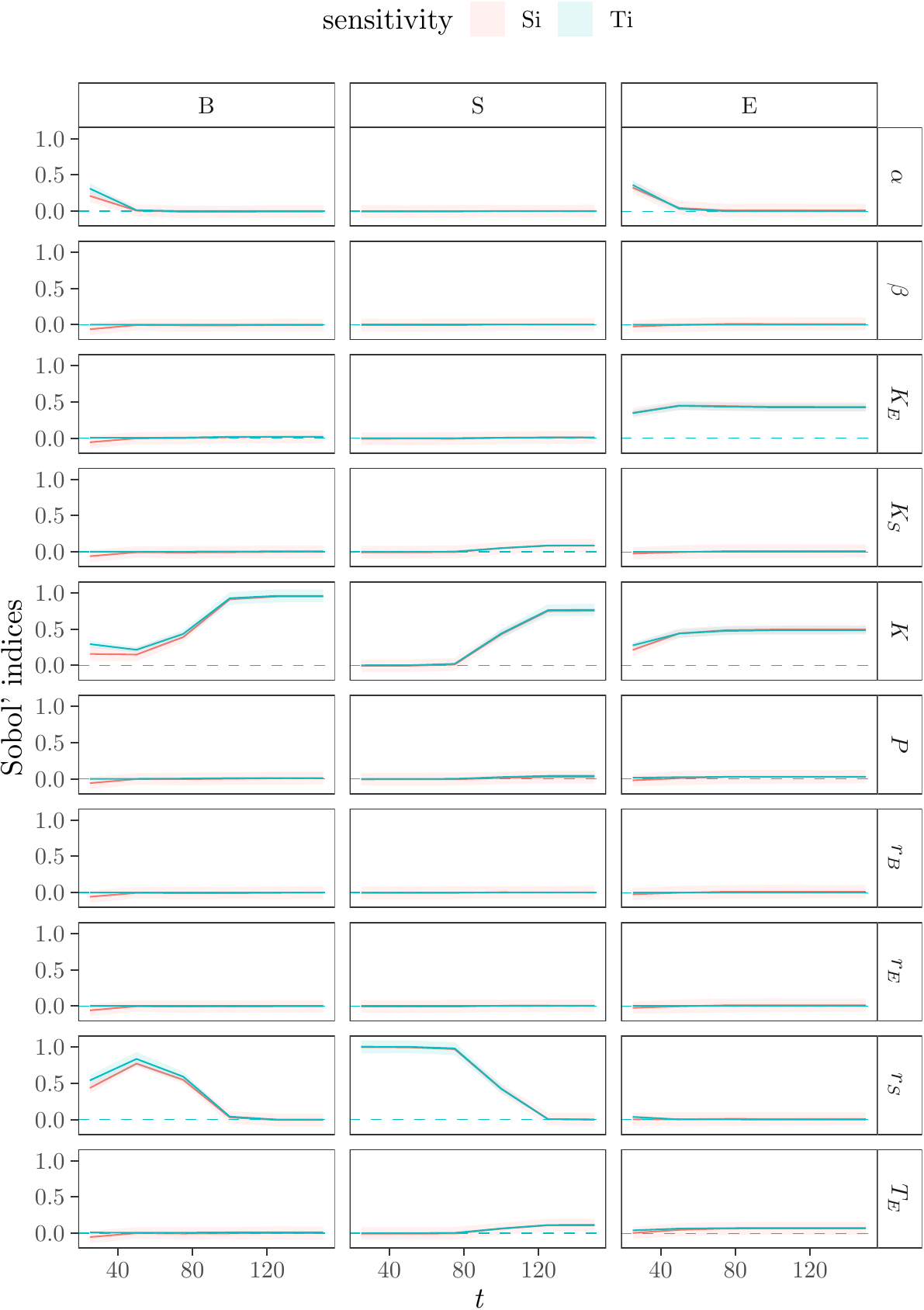} 

}

\caption[Evolution of Sobol' indices through time in the spruce budworm and forest model]{Evolution of Sobol' indices through time in the spruce budworm and forest model. The dashed, horizontal blue line shows the $T_i$ of the dummy parameter.}\label{fig:plot_sobol_budworm_t}
\end{figure}
\end{CodeChunk}

The main results in Figure~\ref{fig:plot_sobol_budworm_t} can be summarized as follows:

\begin{enumerate}
\item The spruce budworm and forest model is largely additive up to $t\approx 80$ as interactions are very small and only affect the behavior of $B$ ($S_\alpha < T_\alpha$, $S_K < T_K$, $S_{r_s} < T_{r_s}$). From $t>80$, the model seems to be fully additive on all three state variables ($S_i \approx T_i$).

\item Given the uncertainty ranges defined in Table~\ref{tab:parameters_budworm}, only four parameters out of 10 are influential in conveying uncertainty to $B$, $S$ and $E$: $\alpha$, $K$, $K_E$ and $r_S$:

\begin{enumerate}
\item The uncertainty in $B$ is determined by $\alpha$, $K$ and $r_S$ at $0<t<100$ and by $K$ at $t>100$.

\item The uncertainty in $S$ is fully driven by $r_S$ up to $t \approx 80$ and by $K$ at $t>120$.

\item The uncertainty in $E$ is influenced by $\alpha$, $K_E$ and $K$ at $0<t<40$ and by $K_E$ and $K$ at $t>40$.
\end{enumerate}

The rest are non-influential and can be fixed at any value without modifying the model output. We should stress here that what is considered ``influential'' in a variance decomposition framework does not necesarily need to concur with the definition of ``influential'' from a systems dynamics perspective.  The influence of a parameter in a variance decomposition framework is determined both by the functional form of the model and the probability distribution selected to describe the uncertainty of the parameters in the model input space.
\end{enumerate}

\hypertarget{conclusions}{%
\section{Conclusions}\label{conclusions}}

\label{sec:conclusions} Mathematical models are used to gain insights
into complex processes, to predict the outcome of a variable of interest
or the explore ``what if'' scenarios. In order to increase their
transparency and ensure the quality of model-based inferences, it is
paramount to scrutinize these models with a global sensitivity analysis.
\pkg{sensobol} aims at furthering the uptake of global sensitivity
analysis methods by the modeling community with a set of functions to
compute variance-based analysis. \pkg{sensobol} allows the user to
combine several first and total-order estimators, to estimate up to
third-order effects and to visualize the results in publication-ready
plots. Due to its integration with \pkg{data.table} and \pkg{deSolve},
\pkg{sensobol} can compute variance-based indices for models with a
scalar or multivariate model output, as well as for systems of
differential equations.

\pkg{sensobol} will keep on developing as the search for more efficient
variance-based estimators is an active field of research. We encourage
the users to provide feedback and suggestions on how can the package be
improved. The most recent updates can be followed on
\url{https://github.com/arnaldpuy/sensobol}.

\hypertarget{acknowledgements}{%
\section{Acknowledgements}\label{acknowledgements}}

This work has been funded by the European Commission (Marie
Sk\l{}odowska-Curie Global Fellowship, grant number 792178 to AP).

\hypertarget{annex}{%
\section{Annex}\label{annex}}

\subsection{Benchmark of sensobol and sensitivity functions}
\label{sec:benchmark}

We compare the execution time of \pkg{sensobol} and \pkg{sensitivity},
from the design of the sample matrix to the computation of the model
output and the Sobol' indices. To ensure that the results are not
critically conditioned by a particular benchmark design, we draw on
\cite{Becker2020} and \cite{Puyj} and compare the efficiency of
\pkg{sensobol} and \pkg{sensitivity} on several randomly defined
sensitivity settings. These settings are created by treating the base
sample size \(N\), the model dimensionality \(k\) and the functional
test of the model as random parameters: \(N\) and \(k\) are described
with the probability distributions shown in Table~\ref{tab:parameters_benchmark} and the test function is
\cite{Becker2020}'s metafunction (Table~\ref{tab:test_functions}, Nº 6),
which randomly combines \(p\) univariate functions in a multivariate
function of dimension \(k\). Becker's metafunction can be called in
\pkg{sensobol} with \code{metafunction()} and its current implementation
includes cubic, discontinuous, exponential, inverse, linear, no-effect,
non-monotonic, periodic, quadratic and trigonometric functions. We
direct the reader to \cite{Becker2020} and \cite{Puyj} for further
information.

\begin{table*}[ht]
\centering
\begin{tabular}{llc}
\toprule
Parameter & Description & Distribution \\
\midrule
$N$ & Base sample size of the sample matrix &  $\mathcal{DU}(10, 100)$\\
$k$ & Model dimensionality & $\mathcal{DU}(3, 100)$\\
\bottomrule
\end{tabular}
\caption{Summary of the benchmark parameters $N$ and $k$. $\mathcal{DU}$ stands for discrete uniform.}
\label{tab:parameters_benchmark}
\end{table*}

We benchmark \pkg{sensobol} and \pkg{sensitivity} as follows:

\begin{itemize}[noitemsep]
\item We create a $(2^{11}, 2)$ sample matrix using random numbers, where the first column is labeled $N$ and the second column $k$.
\item We describe $N$ and $k$ with the probability distributions in Table~\ref{tab:parameters_benchmark}.
\item For $v=1,2,\hdots,2^{11}$ rows, we conducte two parallel sensitivity analysis using the functions and guidelines of \pkg{sensitivity} and \pkg{sensobol} respectively, with the base sample matrix as defined by $N_v$ and $k_v$. The metafunction runs separately in both sensitivity analyses and we bootstrap the sensitivity indices 100 times.
\item We time the computation for both \pkg{sensobol} and \pkg{sensitivity} in each row.
\end{itemize}

The results suggest that \code{sensobol} may be a median of two times
faster than \code{sensitivity}. We provide the code below:

Load required packages:

\begin{CodeChunk}
\begin{CodeInput}
R> library("microbenchmark")
\end{CodeInput}
\end{CodeChunk}

Define the settings of the analysis:

\begin{CodeChunk}
\begin{CodeInput}
R> N <- 2 ^ 11
R> parameters <- c("N", "k")
R> R <- 10 ^ 2
\end{CodeInput}
\end{CodeChunk}

Create the sample matrix:

\begin{CodeChunk}
\begin{CodeInput}
R> dt <- sobol_matrices(matrices = "A", N = N, params = parameters)
R> dt[, 1] <- floor(qunif(dt[, 1], 10, 10 ^ 2 + 1))
R> dt[, 2] <- floor(qunif(dt[, 2], 3, 100))
\end{CodeInput}
\end{CodeChunk}

Run benchmark in parallel:

\begin{CodeChunk}
\begin{CodeInput}
R> n.cores <- makeCluster(floor(detectCores() * 0.75))
R> registerDoParallel(n.cores)
R> y <- foreach(i = 1:nrow(dt),
+              .packages = c("sensobol", "sensitivity")) %dopar%
+   {
+   params <- paste("x", 1:dt[i, "k"], sep = "")
+   N <- dt[i, "N"]
+   out <- microbenchmark::microbenchmark(
+   "sensobol" = {
+   params <- paste("X", 1:length(params), sep = "")
+   mat <- sensobol::sobol_matrices(N = N, params = params, type = "R")
+   y <- sensobol::metafunction(mat)
+   ind <- sensobol::sobol_indices(Y = y, N = N, params = params,
+         first = "jansen", total = "jansen", boot = TRUE, R = R)$results},
+   "sensitivity" = {
+   X1 <- data.frame(matrix(runif(length(params) * N), nrow = N))
+   X2 <- data.frame(matrix(runif(length(params) * N), nrow = N))
+   x <- sensitivity::soboljansen(model = sensobol::metafunction, 
+         X1, X2, nboot = R)
+   },
+   times = 1)
+   }
R> stopCluster(n.cores)
\end{CodeInput}
\end{CodeChunk}

Arrange the data and transform from nanoseconds to milliseconds:

\begin{CodeChunk}
\begin{CodeInput}
R> out <- rbindlist(y)[, time := time / 1e+06]
\end{CodeInput}
\end{CodeChunk}

Plot the results:

\begin{CodeChunk}
\begin{figure}

{\centering \includegraphics{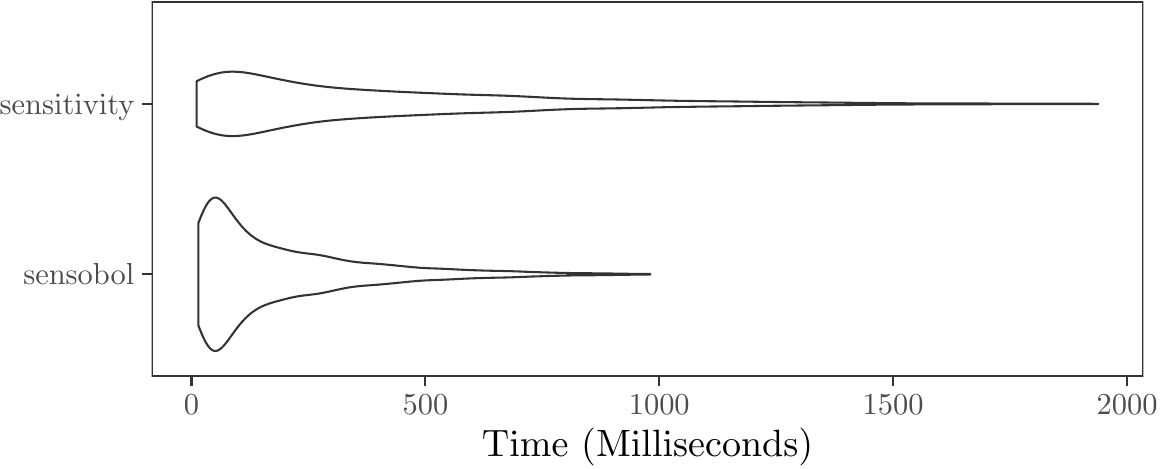} 

}

\caption[Benchmark of the sensitivity and sensobol packages]{Benchmark of the sensitivity and sensobol packages. The comparison has been done with the Jansen estimators.}\label{fig:plot_benchmark}
\end{figure}
\end{CodeChunk}

And compute the median:

\begin{CodeChunk}
\begin{CodeInput}
R> out[, median(time), expr]
\end{CodeInput}
\begin{CodeOutput}
          expr       V1
1:    sensobol 125.6975
2: sensitivity 266.3836
\end{CodeOutput}
\end{CodeChunk}

\subsection{Variogram Analysis of Response Surfaces (VARS-TO)}
\label{sec:VARS}

Given its reliance on variance and co-variance matrices, \pkg{sensobol}
also offers support to compute the variogram analysis of response
surfaces total-order index (VARS-TO) \citep{Razavi2016b, Razavi2016a}.
VARS uses variograms and co-variograms to characterize the spatial
structure and variability of a given model output across the input
space, and allows to differentiate sensitivities as a function of scale
\(h\): if \(\bm{x}_A\) and \(\bm{x}_B\) are two points separated by a
distance \(\bm{h}\), and \(y(\bm{x}_A)\) and \(y(\bm{x}_B)\) is the
corresponding model output \(y\), the variogram \(\gamma(.)\) is
calculated as

\begin{equation}
\gamma(\bm{x}_A-\bm{x}_B) = \frac{1}{2}V \left [y(\bm{x}_A) - y(\bm{x}_B) \right ]\,,
\end{equation}

and the covariogram \(C(.)\) as

\begin{equation}
C(\bm{x}_A-\bm{x}_B) = COV \left [y(\bm{x}_A),  y(\bm{x}_B) \right ]\,.
\end{equation}

Since

\begin{equation}
V \left [y(\bm{x}_A) - y(\bm{x}_B) \right ] = V \left [y(\bm{x}_A) \right ] + V \left [y(\bm{x}_B) \right ] - 2COV \left [ y(\bm{x}_A), y(\bm{x}_B) \right ]\,,
\end{equation}

and \(V \left [ y(\bm{x}_A) \right ] = V \left [ y(\bm{x}_B) \right ]\),
then

\begin{equation}
\gamma (\bm{x}_A - \bm{x}_B) = V \left [ y(\bm{x}) \right ] - C(\bm{x}_A, \bm{x}_B)\,.
\label{eq:variogram}
\end{equation}

If we want to compute the variogram for factor \(x_i\), then

\begin{equation}
\gamma(h_i) = \frac{1}{2}E(y(x_1,...,x_{i+1} + h_i,...,x_n) - y(x_1,...,x_i,...,x_n))^2\,.
\label{eq:dir_variogram}
\end{equation}

Note that the difference in parentheses in Equation~\ref{eq:dir_variogram} involves taking a step along the \(x_i\)
direction and is analogous to computing the total-order index \(T_i\)
(see Section \ref{sec:estimators}). The equivalent of Equation~\ref{eq:variogram} for the model input \(x_i\) would be

\begin{equation}
\gamma_{\bm{x}_{\sim i}^*}(h_i)=V(y|\bm{x}_{\sim i}^*)-C_{\bm{x}_{\sim i}^*}(h_i)\,,
\label{eq:uni_variogram}
\end{equation}

where \(\bm{x}_{\sim i}^*\) is a fixed point in the space of
non-\(x_i\). To compute \(T_i\) in the framework of VARS (labelled as
VARS-TO by \cite{Razavi2016a}), the mean value across the factors' space
should be taken on both sides of Equation~\ref{eq:uni_variogram}, e.g.,

\begin{equation}
E_{\bm{x}_{\sim i}^*} \left [ \gamma_{\bm{x}_{\sim i}}^*(h_i) \right ]=E_{\bm{x}_{\sim i}^*}  \left [ V(y|\bm{x}_{\sim i}^*) \right ] -E_{\bm{x}_{\sim i}^*}  \left [ C_{\bm{x}_{\sim i}}^*(h_i) \right ]\,,
\end{equation}

which can also be written as

\begin{equation}
E_{\bm{x}_{\sim i}^*} \left [ \gamma_{\bm{x}_{\sim i}}^*(h_i) \right ]=V(y)T_i -E_{\bm{x}_{\sim i}^*}  \left [ C_{\bm{x}_{\sim i}}^*(h_i) \right ]\,,
\end{equation}

and therefore

\begin{equation}
T_i=\mbox{VARS-TO}=\frac{E_{\bm{x}_{\sim i}}^*\left [ \gamma_{\bm{x}_{\sim i}}^*(h_i)\right] + E_{\bm{x}_{\sim i}}^* \left [ C_{\bm{x}_{\sim i}}^*(h_i) \right ] }{V(y)}\,.
\label{eq:SM_VARS_ti}
\end{equation}

The computation of VARS does not require \(\bm{A}\),\(\bm{B}\),
\(\bm{A}_B^{(i)}\hdots\) matrices, but a sampling design based on stars.
Such stars are created as follows: firstly, \(N_{star}\) points across
the factor space need to be selected by the analyst using random or
quasi-random numbers. These are the \emph{star centres} and their
location can be denoted as
\(\bm{s}_v = s_{v_1},...,s_{v_i}, ..., s_{v_k}\), where
\(v=1,2,...,N_{star}\). Then, for each star centre, a cross section of
equally spaced points \(\Delta h\) apart needs to be generated for each
of the \(k\) factors, including and passing through the star centre. The
cross section is produced by fixing \(\bm{s}_{v_{\sim i}}\) and varying
\(s_i\). Finally, for each factor all pairs of points with \(h\) values
of \(\Delta h, 2\Delta h, 3\Delta h\) and so on should be extracted. The
total computational cost of this design is
\(N_t=N_{star} \left [ k (\frac{1}{\Delta h} - 1) + 1 \right ]\).

In order to use VARS in \pkg{sensobol}, the analyst should follow the
same steps as in the previous examples. Firstly, she should define the
setting of the analysis, i.e., the number of star centers and distance
\(h\), and create a vector with the name of the parameters:

\begin{CodeChunk}
\begin{CodeInput}
R> star.centers <- 100
R> h <- 0.1
R> params <- paste("X", 1:8, sep = "")
\end{CodeInput}
\end{CodeChunk}

The function \code{vars_matrices()} creates the sample matrix needed to
compute VARS-TO:

\begin{CodeChunk}
\begin{CodeInput}
R> mat <- vars_matrices(star.centers = star.centers, h = h, params = params)
\end{CodeInput}
\end{CodeChunk}

We can then run the model rowwise, in this case the \cite{Sobol1998} G
function (Table~\ref{tab:test_functions}):

\begin{CodeChunk}
\begin{CodeInput}
R> y <- sobol_Fun(mat)
\end{CodeInput}
\end{CodeChunk}

And compute VARS-TO with the \code{vars_to()} function:

\begin{CodeChunk}
\begin{CodeInput}
R> ind <- vars_to(Y = y, star.centers = star.centers, params = params, 
+   h = h)
R> ind
\end{CodeInput}
\begin{CodeOutput}

Number of star centers: 100 | h: 0.1 

Total number of model runs: 7300 
             Ti parameters
1: 0.8213028904         X1
2: 0.2526291054         X2
3: 0.0346579957         X3
4: 0.0104013502         X4
5: 0.0001079858         X5
6: 0.0001034112         X6
7: 0.0001076416         X7
8: 0.0001055614         X8
\end{CodeOutput}
\end{CodeChunk}

The current implementation of \code{vars_to()} does not allow to
bootstrap the indices. This is planned for future \pkg{sensobol}
releases.

\bibliography{jss4353.bib}

\end{document}